# The connected brain: Causality, models and intrinsic dynamics

Adeel Razi[†+] and Karl J. Friston[†]

[†]*The Wellcome Trust Centre for Neuroimaging, University College London, 12 Queen Square, London WC1N 3BG.*

[+]*Department of Electronic Engineering, NED University of Engineering and Technology, Karachi, Pakistan.*

adeel.razi@ieee.org

*Abstract* – Recently, there have been several concerted international efforts – the BRAIN initiative, European Human Brain Project and the Human Connectome Project, to name a few – that hope to revolutionize our understanding of the connected brain. Over the past two decades, functional neuroimaging has emerged as the predominant technique in systems neuroscience. This is foreshadowed by an ever increasing number of publications on functional connectivity, causal modeling, connectomics, and multivariate analyses of distributed patterns of brain responses. In this article, we summarize pedagogically the (deep) history of brain mapping. We will highlight the theoretical advances made in the (dynamic) causal modelling of brain function – that may have escaped the wider audience of this article – and provide a brief overview of recent developments and interesting clinical applications. We hope that this article will engage the signal processing community by showcasing the inherently multidisciplinary nature of this important topic and the intriguing questions that are being addressed.

*Index terms* – dynamic causal modelling · effective connectivity · functional connectivity · resting state · fMRI · graph · Bayesian · intrinsic dynamics

# I. INTRODUCTION

In this review, we will use several key dichotomies to describe the evolution and emergence of modelling techniques used to characterize brain connectivity. Our review comprises three sections. We begin with an historical overview of the brain connectivity literature – starting with the fundamental distinction between functional *segregation* and *integration*. In so doing, we introduce a key difference between *functional* and *effective* connectivity – and emphasize their relationship via underlying models of distributed processing. In the second section, we consider various causal modelling techniques that are used to infer directed brain connectivity. With the help of a unified framework – based on (neuronal) state-space models – we show how (with a succession of simplifying approximations) standard models of connectivity can be derived, and how various measures of statistical dependencies arise from a generative (state-space) model of neuronal dynamics. In the last section, we focus on the application of dynamic causal modelling to endogenous neuronal activity and simulations of neuronal fluctuations based upon the connectome. This section describes a series of recent (and rapid) developments in modelling distributed neuronal fluctuations – and how this modelling rests upon functional connectivity. We will try to contextualize these developments in terms of some historical distinctions above that have shaped our approaches to connectivity in functional neuroimaging.

*Notation:* We use lowercase italics, $x$, for scalars and lowercase bold letters for vectors **x** and vector functions, $\mathbf{x}(t)$, where each element represents a time-dependent state. Matrices are shown as uppercase bold, **X**. In this paper, $*$ corresponds to a convolution operator, † denotes the complex conjugate transpose, $\langle \cdot \rangle$ denotes expectation and ~ denotes discrete time lagged variables. Fourier transforms of variables are in (italic) uppercase such that $\mathbf{FT}(\mathbf{x}(t)) = \mathbf{X}(\omega)$. We use F($\cdot$) to denote a variational free energy functional.

# II. AN HISTORICAL PERSPECTIVE ON BRAIN CONNECTIVITY

The notion of connectivity has long history in brain imaging, which can be traced back to the debates around classicism, modularity and connectionism. In the recent past, a common notion among neuroscientists was that many of the brains functions were predetermined by its structure and its structure was programmed by our genes. This view emphasized functional segregation and localizationism; tracing its history back to the days of phrenology (from Gall in 18th century). Functional localization implies that a function can be localized in a cortical area. This is more general than functional segregation, which suggests that a cortical area is specialized for some aspect of neural processing and that this specialization is anatomically segregated within the cortex. This is similar to an understanding of how computers work; where each part has pre-assigned function that cannot be substituted for by other parts. However, in the past decades this view has changed, with clear evidence that the neural pathways in our brain are flexible, adaptable and connected; able to be molded by changes in our environment or by injury or disease. In short, the brain is quintessentially plastic and can adapt and adopt new functionalities through necessity. This understanding rests on the notion of connectionism (a term first coined by Donald Hebb in 1940's), with the central idea that brain function can be understood as the interaction among simple units; for example, neurons connected by synapses which give rise to a connected whole that changes over time. Connectionism is closely related to (hierarchical) distributed processing, a perspective that has been substantiated by the work of Hubel and Wiesel (Nobel prize in 1981) on how information is processed in visual cortex. They found that visual system comprises *simple* and *complex* cells arranged in hierarchical fashion. This sort of finding underwrites the focus on neural network implementations based on hierarchical distributed constructs – leading to recent exciting developments in machine learning (e.g., hierarchical Bayesian inference[1] and deep learning algorithms [2]).

These ideas emerged in functional brain imaging as functional *segregation* and functional *integration*. Since their inception, there has been a sustained trend to move from functional segregation (and the study of regionally specific brain activation) towards functional integration (and the study of its connectivity). Functional localization implies that a function can be localized to a cortical area, whereas segregation suggests that a cortical area is specialized for some aspects of perceptual or motor processing, and that this specialization is anatomically segregated within the cortex. The cortical infrastructure supporting a single function may then involve many specialized areas whose union is mediated by the functional integration among them. In this view, functional segregation is only meaningful in the context of functional integration and *vice versa*. There are several descriptions of neuronal processing that accommodate the tendency for brain regions to engage in specialized functions (i.e., segregation) and the tendency to coordinate multiple functions (i.e., integration) through coupling specialized regions. This functional integration is a dynamic, self-assembling process, with parts of the brain engaging and disengaging over time – that has been described by appealing to dynamical systems theory; for example, self-organized criticality[3], pattern formation and *metastability* [4].

This review pursues another key theme; namely, the distinction between *functional* and *effective* connectivity. This dichotomy relies on the definition of connectivity (i.e. functional integration) *per se*. The former uses a pragmatic definition of connectivity, based on (Pearson) correlations and rests on statistical dependencies between remote neurophysiological events. But this approach is problematic when dealing with distributed neuronal processes in the brain that are mediated by slender (axonal) neuronal connections or wires. A more mechanistic explanation of observed responses comes from the definition of effective connectivity that refers explicitly to the influence that one neural system exerts over another. [5] proposed that "effective connectivity should be understood as the experiment and time-dependent, simplest possible circuit diagram that would replicate the observed timing relationships between the recorded neurons". This speaks to two important points: effective connectivity is dynamic (activity-dependent), and depends on a model of directed interactions or coupling – which we will focus on in this review. Having said this, an interesting development in functional connectivity now considers temporal dynamics – referred to as dynamic functional connectivity [6]. However, these developments fall short of furnishing a causal explanation of the sort provided by (model-based) effective connectivity. This is because functional connectivity is essentially a description of second order data features, which precludes a mechanistic explanation of neurophysiological timeseries. Recent applications of dynamic causal modelling to ongoing (seizure) activity – in epilepsy – rest explicitly on dynamic functional connectivity to estimate the underlying fluctuations in effective connectivity or cortical gain control [7, 8]. In summary, the operational distinction between functional and effective connectivity is important because it determines the nature of the inferences made about functional integration and the sorts of questions that can be addressed with a careful consideration of intricate interrelationship between effective and functional connectivity [9, 10].

Put simply, functional connectivity is a measure of statistical dependencies, such as correlations, coherence, or transfer entropy. Conversely, effective connectivity corresponds to the parameter of a model that tries to explain observed dependencies (functional connectivity). In this sense, effective connectivity corresponds to the intuitive notion of directed causal influence. This model-based aspect is crucial because it means that the analysis of effective connectivity can be reduced to model comparison; for example, the comparison of a model with and without a particular connection to infer its contribution to observed functional connectivity. In this sense, the analysis of effective connectivity recapitulates the scientific process, because each model corresponds to an alternative hypothesis about how observed data were caused. In our context, these hypotheses pertain to causal models of distributed brain responses. Later, we will consider analytical expressions that link effective and functional connectivity and show that the latter can be derived from the former, whereas the converse is not true.

In summary, we have considered the distinction between functional segregation and integration in the brain and how the differences between functional and effective connectivity shape the way we characterize connections – and the sorts of questions that are addressed to empirical data. In the next section, we look at the relationship between functional and effective connectivity and expand on the causal aspect of effective connectivity. Interested readers are directed to our previous review [10] on brain connectivity for a more detailed discussions.

### III. CAUSAL ANALYSES OF DYNAMICAL SYSTEMS

The brain is a dynamic and self-organizing organ with an emergence dynamics. This dynamics can be seen at multiple spatial and temporal scales; for example, there are tens of thousands of synaptic connections to a single neuron, which can fire dozens of times every second. Furthermore, this connectivity itself changes over multiple spatial and temporal scales. The spatial scale we are interested in, as measured by fMRI, is the macroscopic level; where we are interested in distributed processing or connectivity among neural systems where each neural region or source comprises millions of neurons. As we have previously noted, the direction of information transfer or directed coupling is important. Figure 1 illustrates the fact that changes in connectivity over time underlie the causal relationship among neuronal systems. In Figure 1, we show a graph with *undirected* edges among 10 nodes, where each node can be regarded as a proxy for a neuronal system (in general these nodes could also be network devices in a communication network – exchanging say emails). Alternatively, if the links represent a distance metric, and nodes represent organisms, this could represent a model of how infections are disseminated). In this example, the graph evolves over time. Although the edges of the graph are undirected at each instance, adding a temporal aspect to this evolving graph enables one to infer directed information[1] flow. For example, if we were interested in causal coupling between nodes 1 and 2 (shown as red in Figure 1). We see that the activity in node 1 affects the activity in node 2, where we will assume this influence endures over time. As we see node 1 is connected to node 2 via intermediate nodes: 4, 8, and 5 (shown as blue edges) at time $t - \delta t$; 9, 10 and 7 at time $t$; and 3 at time $t + \delta t$. This means, node 1 can affect node 2 in the future. However, the converse is not true, in that the activity in node 2 cannot affect the future of node 1. This asymmetry is a signature of causality (i.e., temporal precedence) and rests on accounting for the arrow of time. This is why, as we will see below, the statistical models used for characterizing effective connectivity are usually based on *differential* equations (or difference equations in discrete time), and therefore explicitly take time into account. This simple example emphasizes the importance of temporal fluctuations in connectivity, even in undirected graphs. However, we do not want to give the impression that temporal precedence is necessary to infer causal relationships. Temporal precedence is an important aspect – and many definitions of causation require cause to precede effect [12, 13]; for example, directed functional connectivity measures based on Yule-Walker formulations (vector autoregressive models). However, temporal precedence alone cannot distinguish effective connectivity from spurious dependencies caused by unknown factors. As an example, the barometer falls before the rain but it does not cause the rain. The type of causality that we will be concerned with is based on control theoretic concepts, where the causes (either exogenous experimental inputs or endogenous random neural *fluctuations* or both) produce effects (neural activity) that are observed empirically through hemodynamics as blood oxygen level dependent (BOLD) signal. This form of causality is closely related to probabilistic and graphical framework of causal calculus [14] (See Box 1) – although there is a clear distinction between the two approaches that we will return to later.

---

[1] Although we have not used the word 'information' here in a strictly information theoretic sense but there is a straightforward analogy between electrical impulses in neural systems and the classic communications theory picture of source, channel, and receiver [11].

**Box 1:**

**Simpson – Yule Paradox**

The Simpson – Yule paradox, or simply Simpson's paradox, [15-17] – for more recent discussion see [18-20] – refers to the disconcerting situation in which statistical relationships between variables (say $x$ and y) are reversed or negated by the inclusion of an additional variables ($z$). A famous example of this paradox is when the University of California Berkeley came under investigation in 1975 for gender bias in graduate admissions. The graduate admissions statistics revealed that men applying were more likely to be admitted than women. But when data were analysed for each department separately, the reverse was true; in that no department was statistically significant in favour of men. The resolution of this paradox turned out to be that women applied for more competitive departments – with low success rates – in relation to men who went for less competitive majors with a greater chance of being accepted. The main point is that conclusions based on data are sensitive to the variables we choose to hold constant and that is why the `adjustment problem' is so critical in the analysis of observational studies. Even now, no formal procedure has emerged that tells us whether adjusting for variable $Z$ is appropriate for the given study – setting aside intractable criteria [21] based on counterfactuals [22]. However, Simpson's paradox is easily resolved with causal graphs. A simple graphical procedure provides a general solution to the adjustment problem [23]. This procedure is shown in Figure 2 and summarized as follows:

**Objective**: Check if $z_1$ and $z_2$ are sufficient measurements.

*1) $z_1$ and $z_2$ should not be descendants of $x$*

*2) Delete all non-ancestors of { $x, y, z$ }*

*3) Delete all edges from $x$*

*4) Connect any two parents sharing a child*

*5) Strip arrow-heads from all edges*

*6) Delete $z_1$ and $z_2$. Check if $x$ is disconnected from y in the remaining graph, then $z_1$ and $z_2$ are appropriate measurements.*

[Figure 2 reproduced and redrawn with permission from [23]].

*Figure 2 is placed here (inside the box)*

We will use state-space models to describe the basic concepts here and demonstrate that causality based on temporal precedence can be regarded as a special case of causality based on state-space graphs. In what follows, we will look at several measures of causality in functional neuroimaging literature (which refer largely to fMRI but also hold for other modalities like EEG, MEG and local field potentials). These measures can be cast in terms of a generalization of state-space models, based on stochastic differential equations.

*A. State-space modelling of neuronal dynamics*

The most natural framework for modelling distributed and coupled neural activity is to use state-space models. State-space modelling has its origin in control engineering but the term state-space was first used by Kalman [24] and can be traced back to von Bertalanffy who introduced general systems theory to biology in the 1940's and 50's. We start with a generic description of coupled neuronal dynamics[2] in terms of differential equations of the form

$$\dot{\mathbf{x}} = f(\mathbf{x}(t), \boldsymbol{\theta}, \mathbf{u}(t)) + \mathbf{w}(t), \quad \text{(state equation)} \quad (1)$$

$$\mathbf{y}(t) = h(\mathbf{x}(t), \boldsymbol{\theta}) + \mathbf{e}(t), \quad \text{(observation equation)} \quad (2)$$

where $\mathbf{x}(t) = [x_1(t), x_2(t), ... x_n(t)]^T$ represents a vector of $n$ hidden state variables (where each state could correspond to a vast number of neurons in a cortical area, source or spatial mode), $\dot{\mathbf{x}}(t)$ represents the change in those state variables, $\boldsymbol{\theta}$ are the underlying (connectivity) parameters which are assumed to be time-invariant, $\mathbf{y}(t)$ is the observed BOLD signal, whereas $\mathbf{w}(t)$ (resp. $\mathbf{e}(t)$) are state noise (resp. observation or instrument noise) which makes this differential equation random. The (random) endogenous fluctuations[3] $\mathbf{w}(t)$ on the motion of the hidden neuronal states represent the unknown influences (for example spontaneous *fluctuations*) that can only be modelled probabilistically. The neuronal states are hidden as they cannot be measured directly. The function $f$ defines the motion of the coupled dynamical system that is determined by inputs $\mathbf{u}(t)$, which we consider to be deterministic (but could also have stochastic component) and known. Inputs usually pertain to experimentally controlled variables, such as change in stimuli (a visual cue or an auditory signal) or instructions during an fMRI experiment (we will see later that this exogenous input is absent in 'resting state' fMRI). This description of neuronal dynamics provides a convenient model of causal interactions among neuronal populations, as it describes when and where exogenous experimental input $\mathbf{u}(t)$ perturbs the system and how (hidden) states influence changes in other states. Note that we have assumed that the form of the system dependencies $f$ (and the connectivity parameters $\mathbf{\theta}$) are time invariant, which means that we are assuming that the structural properties of the system will remain fixed over time (i.e., over the length of data acquisition).

---

[2] Strictly speaking, the hidden states include both neuronal and haemodynamic states; however, for simplicity, we will ignore haemodynamic states in this paper.

[3] One of our reviewer rightly pointed out that, in this exposition, we have limited ourselves to additive form of endogenous fluctuations that precludes the more general treatment of state-dependent neuronal fluctuation, of the sort $f(x(t), \boldsymbol{\theta}, \boldsymbol{u}(t), \boldsymbol{w}(t))$, which are used in modelling many complex, volatile systems [25] including the brain [26].

We have not discussed the nature of the state and the observation noise process, which we will consider in the last section. For now, we will assume they possess usual noise properties; i.e., they are independent and identically distributed (i.i.d.). We will describe a more general framework for analytic (non-Markovian) random fluctuations in the last section. A key simplification in this form of modelling is that we have lumped together many microscopic neuronal states to form hidden states $\mathbf{x}(t)$ that are abstract representations of neuronal activity (c.f., a mean field approximation). In reality the equations of motion – and the observer equation – describe very complicated interactions among millions of neurons. The formulation above corresponds to the amplitude of macroscopic variables or *order parameters*[4] summarising the dynamics of large neuronal populations. Essentially, this means that the individual neurons become ordered showing a coordinated dynamic pattern that can be described with the concept of order parameters. This sort of formulation can be motivated by basic principles [27]; for example, the centre manifold theorem [28] and the slaving principle [29, 30] that apply generally to coupled dynamical systems.

---

[4] In statistical physics the order parameter is a variable that indicates in which phase you are in; for example in a phase transition between liquid and gas, the order parameter may be the density.

*B. State-space modelling and effective connectivity*

The state and the observation equations in (1) and (2) are generic representations; hence there are several forms that the mappings or functions $f$ and $h$ can take. In turn, these define the sort of inference that can be made – and the nature of causal relationships that can be identified from these models. We will see in this section that almost all models in neuroimaging can be seen as a special case of these equations.

*Dynamic causal modelling*

Although the application of general state-space models in neuroimaging has been around for decades, the explicit use of state-space models based on differential equations can be traced to [31], who first introduced a nonlinear neural mass model for EEG data. However, the most widely used and comprehensive framework – that uses Bayesian statistics to make model and parameter inference – is dynamic causal modelling (DCM) [32]. DCM, when first introduced used ordinary differential equation (ODE) but was later extended to state-space models based on stochastic and random differential equations[33, 34]. The most widely used DCM is based on Taylor expansion of (1) to its bilinear approximations, namely:

$$\dot{\mathbf{x}}(t) = \left(\mathbf{A} + \sum_{j=0}^{J} \mathbf{B}^j \mathbf{u}_j\right)\mathbf{x}(t) + \mathbf{C}\mathbf{u}(t) + \mathbf{w}(t), \tag{3}$$

where $\mathbf{A} = \frac{\partial f}{\partial \mathbf{x}}, \mathbf{B} = \frac{\partial^2 f}{\partial \mathbf{x} \partial \mathbf{u}}$ and $\mathbf{C} = \frac{\partial f}{\partial \mathbf{u}}$ with $\boldsymbol{\theta}_n = \{\mathbf{A}, \mathbf{B}, \mathbf{C}\}$. The matrix $\mathbf{A}$ is known as the Jacobian (or Laplace-Beltrami operator) describing the behaviour – i.e. the effective connectivity – of the system near its fixed point ($f(x_o) = 0$), in the absence of the fluctuations $\mathbf{w}(t)$ and the modulatory inputs $\mathbf{u}(t)$. The matrices $\mathbf{B}^j$ encode the change in effective connectivity induced by the *j*th input $\mathbf{u}_j(t)$ and $\mathbf{C}$ embodies the strength of the direct influences of inputs $\mathbf{u}(t)$ on neural activity. In fMRI, the mapping from hidden states to the observed BOLD data $y(t)$ is based on a haemodynamic model, which transforms hidden neuronal states of each population or region into predicted BOLD responses – using a previously established biophysical model [32, 35, 36]. This haemodynamic model is based on four ordinary differential equations and five haemodynamic parameters $\boldsymbol{\theta}_h$, such that $\boldsymbol{\theta} = \{\boldsymbol{\theta}_n, \boldsymbol{\theta}_h\}$. The haemodynamic model describes how neuronal activity engenders vasodilatory signals that lead to increases in the blood flow, which in turn changes the blood volume and deoxyhemoglobin content that subtends the measured signal.

The bilinear approximation to our general state-space model of neurophysiological dynamics furnishes a probabilistic model that specifies the probability of observing any time series given the parameters. This is known as a likelihood model and usually assumes the observed data are linear mixture of the model predictions and Gaussian observation noise. By combining this likelihood model with prior beliefs (specified in terms of probability distributions) we have, what is called in Bayesian statistics, a generative model. This allows one to use standard (variational) procedures to estimate the posterior beliefs about the parameters and, crucially, the model itself. Herein lies the real power of dynamic causal modeling; namely, the ability to compare different models of the same data. This comparison rests on the model evidence, which is simply the probability of the observed data, under the model in question (and given known or designed exogenous inputs). The evidence is also called the marginal likelihood because one marginalizes or removes dependencies on the unknown quantities (hidden states and parameters). The model evidence can simply be written as:

$$p(\mathbf{y}|m, \mathbf{u}) = \int p(\mathbf{y}, \mathbf{x}, \boldsymbol{\theta}|m, \mathbf{u}) d\mathbf{x} d\boldsymbol{\theta}. \tag{4}$$

Model comparison rests on the evidence for one model relative to another (see Penny et al. 2004 for a discussion in the context of fMRI). Model comparison based on the likelihood of different models provides the quantitative basis for all evidence-based hypothesis testing. Usually, one selects the best model using Bayesian model comparison, where different models are specified in terms of priors on the coupling parameters. These are used to switch off parameters by assuming *a priori* that they are zero (to create a new model). In DCM, priors used are so-called 'shrinkage-priors', because the posterior estimates shrink towards the prior mean. The size of the prior variance determines the amount of shrinkage. With a null model $m_0$ and an alternate model $m_1$ Bayesian model comparison rests on computing the logarithm of the evidence ratio:

$$\ln\left(\frac{p(\mathbf{y}|m_1)}{p(\mathbf{y}|m_0,)}\right) = \ln p(\mathbf{y}|m_1) - \ln p(\mathbf{y}|m_0)$$

$$\approx F(\mathbf{y}, \boldsymbol{\mu}_1) - F(\mathbf{y}, \boldsymbol{\mu}_0). \tag{5}$$

where F(.) is the free energy which provides an (upper bound) approximation to Bayesian model evidence. Note that we have expressed the logarithm of the marginal likelihood ratio as a difference in log-evidences. This is the preferred form, because model comparison is not limited to two models, but can cover a large number of models, whose quality can be usefully quantified in terms of their log-evidences. A relative log-evidence of three corresponds to a marginal likelihood ratio (Bayes factor) of about 20 to 1, which is considered strong evidence in favor of one model over another [37]. An important aspect of model evidence is that it includes a complexity cost (which is not only sensitive to the number of parameters but also to their interdependence). This means that a model with redundant parameters would have less evidence, even though it provided a better fit to the data (see Penny et al. 2004). In most current implementations of dynamic causal modeling, the log-evidence is approximated with a (variational) free-energy bound that (by construction) is always less than the log-evidence. As we see in (5), this bound is a function of the data and (under Gaussian assumptions about the posterior density) some proposed values for the states and parameters. When the free-energy is maximized (using gradient ascent) with respect to the proposed values, they become the maximum posterior or conditional estimates, $\boldsymbol{\mu}$ and the free-energy, $F(\mathbf{y}, \boldsymbol{\mu}_1) \leq \ln p(\mathbf{y}|m)$ approaches the log-evidence. We will return later to Bayesian model comparison and inversion of dynamic causal models. At the moment, we consider some alternative models. The first is a discrete-time linear approximation to (1), which is the basis of Granger causality.

*Vector autoregressive modelling*

In contrast to DCM – where causality is based on control theoretic constructs, (multivariate) autoregressive models [38-40] use temporal precedence for inferring causality in BOLD time series [41]. This is known as directed functional connectivity in neuroscience. It is straightforward to see that one can convert a state-space model – or DCM – into a vector autoregressive model, with a few simplifying assumptions. Using a linear approximation to the state-space model of (1) and assuming that we can measure the neuronal states directly (i.e., $\mathbf{y}(t) = \mathbf{x}(t)$), then we can write

$$\mathbf{y}(t) = \widetilde{\mathbf{A}}\mathbf{x}(t - \delta) + \mathbf{z}(t), \tag{6}$$

which can be written as

$$\mathbf{Y} = \widetilde{\mathbf{Y}}\widetilde{\mathbf{A}}^T + \mathbf{Z},$$

where $\widetilde{\mathbf{A}} = \exp(\delta \mathbf{A})$ and $\mathbf{z}(t) = \int_0^\delta \exp(\tau \mathbf{A}) w(t-\tau) d\tau$. The second equality expresses the resulting vector autoregression model as a simple general linear model, with explanatory variables, $\widetilde{\mathbf{Y}}$ that correspond to a time-lagged (time x source) matrix of states. Here, the unknown parameters comprise the autoregression matrix $\widetilde{\mathbf{A}}$. Note that the *innovations*, $\mathbf{z}(t)$ are now a mixture of past fluctuations in $\mathbf{w}(t)$ that are remembered by the system. There is clear distinction between fluctuations $\mathbf{w}(t)$ that drives the hidden states (1) compared to the innovations $\mathbf{z}(t)$ in (6) that underlie autoregressive dependencies among observation $\mathbf{y}(t)$. There is an important point to note here. Because the re-parameterisation of the effective connectivity in (3) uses a matrix exponential, the autoregressive coefficients $\widetilde{\mathbf{A}}$ in (6) are no longer the parameters of the underlying effective connectivity among neuronal states. This means that any model comparisons – based on classical likelihood ratio tests such as BIC – will be making inferences about the statistical dependencies modelled by the autoregressive process and not about the causal coupling as in DCM. This is why connectivity measures based on autoregressive coefficients – e.g., Granger causality [42] – are regarded as directed functional connectivity as opposed to effective connectivity. A further distinction is that most Granger causality applications either ignore haemodynamic convolution, or assume that haemodynamics are identical and noiseless (David et al. 2008). An important aspect of Granger causality measures based on autoregressive formulations (we provide analytic links between the two below) is that they can become unreliable in the presence of measurement noise and more so when underlying dynamics is dominated by slow (unstable) modes – quantified by the principal *Lyapunov exponent* [43]. However, there are several recent advances in the Granger causality literature which speak to these limitations [44-46].

*Structural Equation Modelling*

Structural equation modelling (SEM) [47] is another generic approach developed primarily in economics and social sciences [48, 49] and used in the (structural) neuroimaging for the first time in [50]. We can again see that SEM is a special case of (1) by appealing to the (adiabatic) assumption that neuronal dynamics have reached equilibrium at each point of observation – or, in other words, the dynamics are assumed to occur over a time scale that is short relative to the fMRI sampling interval. In terms of implementation, we can force this condition by having very strong shrinkage priors in DCM. With this assumption we can reduce the generative model of (3) so that it predicts the observed covariance among regional response over time instead of predicting the time series itself. Mathematically, this means that we assume $\mathbf{y}(t) = \mathbf{x}(t)$ and $\mathbf{u}(t) = 0$ and $\dot{\mathbf{x}}(t) = 0$. This simply means that $\mathbf{x}(t) = \mathbf{y}(t) = -\mathbf{A}^{-1}\mathbf{w}(t)$ which implies that

$$\mathbf{\Sigma}_y = \mathbf{A}^{-1}\mathbf{\Sigma}_w(\mathbf{A}^{-1})^T, \qquad (7)$$

where $\mathbf{\Sigma}_y = \langle \mathbf{y}(t)\mathbf{y}(t)^T \rangle$ and $\mathbf{\Sigma}_w = \langle \mathbf{w}(t)\mathbf{w}(t)^T \rangle$. Note that we do not have to estimate hidden states, because the generative model explains observed covariances in terms of random fluctuations and unknown coupling parameters. The form of (7) has been derived from the generic generative model. In this form, it can be regarded as a Gaussian process model, where the coupling parameters become, effectively, parameters of the covariance among observed signals due to the hidden states. We can also give an alternate formulation of SEM in terms of *path coefficients* but we skip this for brevity – for details see [51].

Although, SEM has been used in fMRI literature, SEM provides a description of static dependencies; hence it is not suitable for fMRI (and EEG/MEG) timeseries, where the characteristic time constants of the neuronal dynamics and haemodynamics are much larger than the exogenous inputs that drive them. This means that testing for context-sensitive changes in effective connectivity becomes problematic in event-related designs. For example, [52], used simulated fMRI timeseries from a realistic network model, for two task conditions, in which the anatomical connectivity is known and can be manipulated. The results suggested that caution is necessary in applying SEM to fMRI data, and illustrate that functional interactions among distal network elements can appear abnormal, even if only part of a network is damaged.

Another issue, when using SEM to infer effective connectivity, is that we can only use models of low complexity – usually, (acyclic) models that have no recurrent connections [53]. This is because fitting the sample covariance means that we have to throw away a lot of information in the original time series. Heuristically, the ensuing loss of degrees of freedom means that conditional dependencies among the estimates of effective connectivity are less easy to resolve. In machine learning literature, structural equation modeling can be regarded as a generalization of inference on linear Gaussian Bayesian networks that relaxes the acyclic constraint. As such, it is a generalization of *structural causal modeling*, which deals with directed acyclic graphics (see next section). This generalization is important in the neurosciences, because of the ubiquitous reciprocal connectivity in the brain that render it cyclic or recursive.

Next, we turn to the description of time series based on second order statistics and show that they can analytically be derived from the state-space model of (1).

*Coherence, cross spectra and correlations*

Hitherto, we have only considered procedures for identifying effective connectivity from fMRI time series. However, an important question remains: is there an analytical relationship between functional and effective connectivity? Figure 3 addresses this question schematically by showing how various measures of statistical dependencies (functional connectivity) are interrelated – and how they can be generated from a dynamic causal model. This schematic contextualises different measures of functional connectivity and how they arise from (state-space) models of effective connectivity. In other words, measures that are typically used to characterize observed data can be regarded as samples from a probability distribution over functions, whose expectation is known. This means that one can treat normalized measures – like cross-correlation functions and spectral Granger causality – as an explicit function of the parameters of the underlying generative process.

In this schematic we have included common (descriptive) measures of functional connectivity that have been used in fMRI. These include the correlation coefficient (the value of the cross correlation function at zero lag), coherence and (Geweke) Granger causality [54]. These measures can be regarded as standardised (second-order) statistics based upon the cross covariance function, the cross spectral density and the directed transfer functions respectively. In turn, these are determined by the first-order (Volterra) kernels, their associated transfer functions and vector autoregression coefficients. For readers not familiar with Volterra kernels, the use of Volterra kernels provides an alternative to the conventional identification methods by expressing the output signal as high-order nonlinear convolution of the inputs.

It can simply be thought of as a functional Taylor expansion and can be regarded as power series with memory (see [55]for detailed discussion).Crucially, all these representations can be generated from the underlying state-space model used by DCM. Let us examine these relationships further. First, there is a distinction between the state-space model (upper two boxes) – that refers to hidden or system states – and representations of dependencies among observations (lower boxes) – that do not. This is important because although one can generate the dependencies among observations from the state-space model, one cannot do the converse. In other words, it is not possible to derive the parameters of the state-space model (e.g., effective connectivity) from transfer functions or autoregression coefficients. This is why one needs a state-space model to estimate effective connectivity or – equivalently – why effective connectivity is necessarily model-based. Second, we have seen in previous sections that SEM and autoregressive representations can be derived from (bilinear and stochastic) DCM in a straightforward manner (under certain assumptions). The convolution kernel representation in Figure 3 provides a crucial link between covariance-based second-order measures – like cross covariance, cross correlation – and their spectral equivalents, like cross spectra and coherence. Figure 3 also highlights the distinction between second order statistics (lower two rows) and models of the variables *per se* (upper three rows). For example, convolution and autoregressive representations can be used to generate timeseries (or their spectral counterparts), while cross covariance functions and autoregression coefficients describe their second-order behaviour. This is important because this second-order behaviour can be evaluated directly from observed timeseries. Indeed, this is the common way of measuring functional connectivity in terms of (second-order) statistical dependencies.  We also highlight the dichotomy between time and frequency representations (measures within the light Gray box). For example, the (first-order Volterra) kernels in the convolution formulation are the Fourier transform of the transfer functions in frequency space (and *vice versa*). Similarly, the directed transfer functions of the autoregressive formulation are based upon the Fourier transforms of the autoregression coefficients. Another distinction is between representations that refer explicitly to random (state and observation) noise and autoregressive representations that do not. For example, notice that the cross-covariance functions of the data depend upon the cross-covariance functions of state and observation noise. Conversely, the autoregression formulation only invokes (unit normal) innovations (although the autoregression coefficients are an implicit function of both state and observation noise covariance functions). In the current setting, autoregressive representations are not regarded as models, but simply ways of representing dependencies among observations. This is because (hemodynamic) responses do not cause responses – hidden (neuronal) states cause responses.

Crucially, all of the above formulations of statistical dependencies contain information about temporal lags (in time) or phase delays (in frequency). This means that, in principle, all measures are directed – in the sense that the dependencies from one region to another are distinct from the dependencies in the other direction. However, only the autoregressive formulation provides directed measures of dependency – in terms of directed transfer functions or Granger causality. This is because the cross-covariance and spectral density functions between two timeseries are antisymmetric. The autoregressive formulation can break this (anti) symmetry because it precludes instantaneous dependencies by conditioning the current response on past responses. Note that Granger causality is – in this setting – a measure of directed functional connectivity [56]. This means that Granger causality (or the underlying autoregression coefficients) reflects directed statistical dependencies – such that two regions can have strong autoregression coefficients or Granger causality in the absence of a direct effective connection. Finally, there is a distinction between (second order) effects sizes in the upper row of dependency measures and their standardised equivalents in the lower row. For example, coherence is simply the amplitude of the cross spectral density normalised by the auto-spectra of the two regions in question. Similarly, one can think of Granger causality as a standardised measure of the directed transfer function (normalised by the auto-spectra of the source region).

We also note another widely used measure of functional dependencies, known as mutual information [57] that quantifies the shared information between two variables – and can reflect both linear and nonlinear dependencies. For example, if two time-series are independent, there is no shared information and hence the mutual information is zero. Mutual information can be calculated relatively simply – under the assumption that time series are Gaussian – from coherence in the frequency domain as [58-60]

$$\vartheta_{ij} = \frac{1}{2\pi} \int_{\omega_1}^{\omega_2} \log\left(1 - C_{ij}(\omega)\right) d\omega, \tag{8}$$

where $C_{ij}(\omega)$ is the coherence – as defined on Figure 3 – between the two time series $i$ and $j$.

In summary, given a state-space model, one can predict or generate the functional connectivity that one would observe, in terms of cross-covariance functions, complex cross-spectra or autoregression coefficients (where the latter can be derived in a straightforward way from the former using the Yule-Walker formulation). In principle, this means that one could either use the sampled cross-covariance functions or cross-spectra as data features. It would also be possible to use the least-squares estimate of the autoregression coefficients – or indeed Granger causality – as data features to estimate the underlying effective connectivity. We will describe such schemes in the next section.

*C. Summary*

This section has tried to place different analyses of connectivity in relation to each other. The most prevalent approaches to effective connectivity are dynamic causal modeling, structural equation modeling, and Granger causality. We have highlighted some of the implicit assumptions made when applying structural equation modeling and Granger causality to fMRI time series. In the remainder of this review, we will focus on generative models of distributed brain responses and consider some of the exciting developments in this field.

IV. BIOPHYSICAL MODELLING OF NEURONAL DYNAMICS

Biophysical models of neuronal dynamics are usually used for one of two things: either to understand the emergent properties of neuronal systems or as observation models for measured neuronal responses. We discuss examples of both. In terms of emergent behaviors, we will consider dynamics on structure [61-69] and how this behavior has been applied to characterizing autonomous or endogenous fluctuations in fMRI [70-73]. This section concludes with recent advances in dynamic causal modeling of directed neuronal interactions that support endogenous fluctuations. Some subsections below are based on our previous review [10].

*A. Intrinsic dynamics, criticality and bifurcations*

The use of resting state fMRI [74, 75] or studies based on BOLD signal correlations while the brain is at rest are widespread [76]. These patterns are thought to reflect anatomical connectivity [77] and can be characterized in terms of remarkably reproducible spatial modes (resting-state or intrinsic networks). One of these modes recapitulates the pattern of deactivations observed across a range of activation studies (the default mode; [78]). Resting state fMRI studies show that even at rest, endogenous brain activity is self-organizing and highly structured. The emerging picture is that endogenous fluctuations are a consequence of dynamics on anatomical connectivity structures with particular scale-invariant characteristics [70, 71, 79, 80].These are well-studied and universal characteristics of complex systems and suggest that we may be able to understand the brain in terms of universal phenomena [81]. Universality is central to the hypothesis that the cerebral cortex is poised near a critical point, where only one variable, a control parameter determines the macroscopic behavior of the system [82, 83]. This is an important issue because systems near phase-transitions show universal phenomena [84-88]. Near the critical point, correlations between neurons would occur across all scales, leading to optimized communication [89]. Experimental evidence for this notion has accumulated over the past decades, where power laws and scaling relationships have been found in human neuroimaging timeseries [90, 91]. However, it should be noted that with more attention on this new direction, there are a variety of distributions; e.g., stretched exponential, Rayleigh, double exponential and lognormal that are found in neurophysiological timeseries [26, 92, 93]. Hence there may be a need to carefully disambiguate the causes of these heavy-tailed distributions found in the brain and behaviour. From the dynamical system perspective, endogenous dynamics are thought to be generated by the dynamic instabilities that occur near *bifurcations;* i.e., dynamics that accompany a loss of stability when certain control parameter(s) reach a critical value [26, 94-96]. The eigenmodes of neuronal (effective) connectivity that define the stability of the resting state gives rise to scale-free fluctuations that emerge from the superposition of the few modes that decay slowly. These slowly fluctuating (unstable) modes have Lyapunov exponents that are close to zero. This occurs when systems approach *transcritical* bifurcations (or stochastic Hopf bifurcations when the eigenvalues are complex [97, 98] and show critical slowing [93]). Put simply, this means that the ensuing networks are defined by trajectories that have fixed points close to instability. This means that the neuronal fluctuations persist over longer time scales to generate the patterns responsible for the emergence of intrinsic brain networks. The amplitudes of these eigenmodes or patterns correspond to the order parameters described in Section III (B). The (negative) inverse of the Lyapunov exponent corresponds to the characteristic time constant of each mode, where each mode with a small exponent (large time constant) corresponds to an intrinsic brain network or resting state networks (RSN).

*B. Causal modelling of neuronal dynamics*

The past decade has seen the introduction of graph theory to brain imaging. Graph theory provides an important formulation for understanding dynamics on structure. Developments in this area have progressed on two fronts: namely, to understand connections between graphs and probability calculus and the use of probabilistic graphs to resolve causal interactions. The probabilistic graph framework goes beyond classical constructs by providing powerful symbolic machinery and notational convenience (e.g., the use of dependency graphs to resolve Simpson's paradox: see Box 1). Within this enterprise one can differentiate at least two streams of work: one based on Bayesian dependency graphs or graphical models called *structural causal modeling* [99], and the other based on causal influences over time, which we consider under *dynamic causal modeling*. Structural causal modeling originated with structural equation modeling [47] and uses graphical models (Bayesian dependency graphs or Bayes nets), in which direct causal links are encoded by directed edges. These tools have been largely developed by Pearl [22] and are closely related to the ideas of [100-102]. An essential part of network discovery in structural causal modeling is the concept of intervention; namely, eliminating connections in the graph and setting certain nodes to given values. Structural causal modelling lends a powerful and easy-to-use graphical method to show that a particular model specification identifies a causal effect of interest. Moreover, the results derived from structural causal modelling do not require specific distributional or functional assumptions like multivariate normality, linear relationships *etc*. However, it is not the most suitable framework to understand coupled dynamical systems, because it is limited in certain respects. Crucially, it only deals with conditional independencies in directed acyclic graphs (DAG). This is problematic because brains perform computations on a directed and cyclic graph. Every brain region is connected reciprocally (at least poly-synaptically), and every computational theory of brain function rests on some form of reciprocal or reentrant message passing. Another drawback is that the causal calculus of structural causal modelling ignores time. Pearl argues that a causal model should rest on functional relationships between variables. However, these functional relationships cannot deal with (cyclic) feedback loops. Pearl [14] argues for dynamic causal models when attempting to identify hysteresis effects, where causal influences depend on the history of the system. Interestingly, the DAG restriction can be finessed by considering dynamics and temporal precedence within structural causal modeling. This is because the arrow of time can be used to convert a directed cyclic graph into an acyclic graph when the nodes are deployed over successive time points. This leads to structural equation modeling with time-lagged data and related autoregression models, such as those employed by Granger causality described above. As established in the previous section, these can be regarded as discrete time formulations of dynamic causal models in continuous time.

*Structural and dynamic causal modeling*

As established above, in relation to the modelling of fMRI timeseries, dynamic causal modeling refers to the (Bayesian) inversion and comparison of models that cause observed data. These models are usually state-space models expressed as (ordinary, stochastic, or random) differential equations that govern the motion of hidden neurophysiological states. These models are generally equipped with an observer function that maps from hidden states to observed signals (see (1)). The basic idea behind DCM is to formulate one or more models of how data are caused in terms of a network of distributed sources. These sources talk to each other through parameterized connections and influence the dynamics of hidden states that are intrinsic to each source. Model inversion provides estimates of their parameters and the model evidence.

We have introduced DCM for fMRI using a simple state-space model based on a bilinear approximation – extensions to for e.g. nonlinear [103] and two state [104] DCM, among others, are also available and are in use – to the underlying equations of motion that couple neuronal states in different brain regions [32]. Most DCMs consider point sources both for fMRI and MEG/EEG data (c.f., equivalent current dipoles) and are formally equivalent to the graphical models used in structural causal modeling. However, in dynamic causal modeling, they are used as explicit generative models of observed responses. Inference on the coupling within and between nodes (brain regions) is generally based on perturbing the system and trying to explain the observed responses by inverting the model. This inversion furnishes posterior or conditional probability distributions over unknown parameters (e.g., effective connectivity) and the model evidence for model comparison [105]. The power of Bayesian model comparison, in the context of dynamic causal modeling, has become increasingly evident. This now represents one of the most important applications of DCM and allows different hypotheses to be tested, where each DCM corresponds to a specific hypothesis about functional brain architectures [106-112]. DCM has been mostly used for (task based) fMRI and electrophysiological dynamics (EEG/MEG/LFPs) but most recent advances have focused on the modelling of intrinsic brain networks in the absence of exogenous influence – known as 'resting state' fMRI [74]. In the remainder of this section, we will briefly review these developments besides discussing these new mathematical models we will also showcase some of their clinical applications to neurodegenerative diseases, like Parkinson's disease.

*C. Dynamic casual modelling of intrinsic networks*

There has been an explosion of research examining spontaneous fluctuations in fMRI signals (see Figure 4). These fluctuations can be attributed to the spontaneous neuronal activity, which is usually ignored in deterministic models of responses to (designed) experimental inputs. Deterministic DCMs are cast as multiple input multiple output (MIMO) systems, where exogenous inputs perturbed the brain to produce an observed BOLD response. In absence of external inputs – as in the case of resting state fMRI – neuronal networks are driven by activity that is internal to the system [113]. The generative model for resting state fMRI timeseries has the same form as (3) but discounts exogenous modulatory input. It is to be noted that we can still include exogenous (or experimental) inputs, $u(t)$ in our model. These inputs drive the hidden states —and are usually set to zero in resting state models. It is perfectly possible to have external, (non-modulatory) stimuli, as in the case of conventional functional neuroimaging studies. For example, in [114] we used an attention to visual motion paradigm to illustrate this point. Figure 5 provides a schematic of the resulting stochastic dynamic causal model. In contrast to the previous section, we will adopt a generalized framework in which state noise $w(t)$ and observation noise $e(t)$ are analytic (i.e., non-Markovian). This simply means that *generalized motion* of the state noise $\mathbf{w}(t) = [\mathbf{w}(t), \dot{\mathbf{w}}(t), \ddot{\mathbf{w}}(t) \dots]$ is well defined in terms of its higher-order statistics. Similarly, the observation noise $\tilde{\mathbf{e}}(t) = [\mathbf{e}(t), \dot{\mathbf{e}}(t), \ddot{\mathbf{e}}(t) \dots]$ has a well-defined covariance (for a more detailed discussion see [115].

Consequently, the stochastic part of the generative model in Equation (1) can be conveniently parameterised in terms of its precision (inverse covariance). This allows us to cast (1) as a random differential equation instead of stochastic differential equation; hence eschewing *Ito calculus* [34, 116]. Interested readers will find a theoretical motivation for using analytic state noise in [34]. Under linearity assumptions, (1) can be written compactly in generalized coordinates of motion:

$$\mathbf{D}\tilde{\mathbf{x}}(t) = \tilde{f}(\tilde{\mathbf{x}}, \tilde{\mathbf{u}}, \boldsymbol{\theta}) + \widetilde{\mathbf{w}}(t), \quad (9)$$
$$\tilde{\mathbf{y}}(t) = \tilde{h}(\tilde{\mathbf{x}}, \boldsymbol{\theta}) + \tilde{\mathbf{e}}(t),$$

where **D** is the block diagonal temporal derivative operator, such that the vectors of generalized coordinates of motion are shifted as we move from lower-orders of motion to higher-orders [115]. For resting state activity, (9) takes a very simple linear form:

$$\mathbf{D}\tilde{\mathbf{x}}(t) = \mathbf{A}\tilde{\mathbf{x}}(t) + \mathbf{C}\tilde{\mathbf{u}}(t) + \tilde{\mathbf{v}}(t). \quad (10)$$

This is an instance of a linear dynamical system with quasi-deterministic behaviour [117, 118]. Put simply, the linear dynamical system described by (10) is insensitive to the initial conditions; hence, it can only exhibit a limited repertoire of behaviour: linear systems can contain closed orbits, but they will not be *isolated*, hence no limit cycles – either stable or unstable – can exist, which precludes chaotic behaviour. Technically speaking, if $\boldsymbol{\lambda}$ represents the eigenvalues of the Jacobian $\partial_{\tilde{x}} f = \mathbf{A}$, that is $\boldsymbol{\lambda} = \mathbf{v}^{\dagger}\mathbf{A}\mathbf{v}$, where † denotes the generalized inverse, then the Lyapunov exponents $\Re(\boldsymbol{\lambda})$ of this linear dynamical system will always be negative. In general, the Jacobian is not symmetrical (causal effects are asymmetric); hence the modes and eigenvalues take complex values. For the detailed treatment of the special case of symmetrical connectivity – in which the eigenmodes of functional and effective connectivity become the same – see [119]. It is also worth noting that these eigenmodes are also closely related to (group) independent component analysis (ICA) except with a rotation based on higher order statistics – for details see [120].

There are currently two schemes to invert models of the form (9). They differ in what data features they use for the parameter estimation. The first inverts data in the time domain and the model is used to predict the timeseries *per se*. This is referred to as stochastic DCM [116]. The second approach makes predictions in the frequency domain and is based on fitting second-order data features like cross-spectra. This is referred to as spectral DCM [114, 121]. We will briefly review both schemes and illustrate their clinical applications. For a schematic illustration of dynamics causal modelling of intrinsic dynamics, please see Figure 6. Figure 7 presents a comparison of the two schemes

*Stochastic dynamic causal models*

Stochastic DCM entails inverting a model of the form given by (10) in the time domain, which includes state noise. This requires estimation of not only the model parameters (and any *hyperparameters* that parameterise the precision of generalised random fluctuations), but also the hidden states, which become random (probabilistic) variables. Hence the unknown quantities to be estimated under a stochastic DCM are $\boldsymbol{\psi} = \{\tilde{\mathbf{x}}(t), \boldsymbol{\theta}, \boldsymbol{\sigma}\}$, where $\boldsymbol{\sigma}$ refers to any hyperparameters describing random fluctuations. In terms of temporal characteristics, the hidden states are time-variant, whereas the model parameters (and hyperparameters) are time-invariant.

There are various variational schemes in literature that can invert such models. For example, dynamic expectation maximization (DEM) [122] and generalized filtering (GF) [34]. There is a subtle but important distinction between DEM and generalised filtering. DEM calls on the mean field approximation described above i.e., it assumes $q(\boldsymbol{\psi}) = q(\tilde{\mathbf{x}}(t))q(\boldsymbol{\theta})q(\boldsymbol{\sigma})$, whereas generalized filtering, as the name suggest, is more general that it does not make this assumption. Both schemes, however, assume a fixed form Gaussian distribution for the approximate conditional posterior densities (the Laplace approximation). Generalized filtering considers all unknown quantities to be conditionally dependant variables i.e., $q(\boldsymbol{\psi}) = q(\tilde{\mathbf{x}}, \boldsymbol{\theta}, \boldsymbol{\sigma})$, and produces time-dependent conditional densities for all unknown quantities. The time-*invariant* parameters and hyperparameters are cast as time-variant with the prior constraint that their temporal variation is small. In brief, this online scheme assimilates log-evidence at each time point, in the form of variational free energy and provides time-dependant conditional densities for all unknown quantities. This is in contrast to schemes, like DEM (or deterministic model inversion using variational Laplace) with mean field approximations that assimilates all the data before computing the free energy.

Figure 8 shows an exemplar data analysis reported in [123] that used stochastic DCM to quantify effective connectivity changes in Parkinson's disease. Depleted of dopamine, the dynamics of the Parkinsonian brain impact on both 'action' and 'resting' motor activity. Deep brain stimulation (DBS) has become an established means of managing these symptoms, although its mechanisms of action remain unclear. Kahan et al. modeled the effective connectivity – using stochastic dynamic causal modelling – underlying low frequency BOLD fluctuations in the resting Parkinsonian motor network. They were particularly interested in the distributed effects of DBS on cortico-subcortical connections. Specifically, they showed (see Figure 8) that subthalamic nucleus (SN) deep brain stimulation modulates all the major components of the motor cortico-striato-thalamo-cortical loop, including the cortico-striatal, thalamo-cortical, direct and indirect basal ganglia pathways, and the hyperdirect subthalamic nucleus projections. The strength of effective subthalamic nucleus afferents and efferents were reduced by stimulation, whereas cortico-striatal, thalamo-cortical and direct pathways were strengthened. Remarkably, regression analysis revealed that the hyperdirect, direct, and basal ganglia afferents to the subthalamic nucleus predicted clinical status and therapeutic response to deep brain stimulation; however, suppression of the sensitivity of the subthalamic nucleus to its hyperdirect afferents by deep brain stimulation may subvert the clinical efficacy of deep brain stimulation. These findings highlight the distributed effects of stimulation on the resting motor network and provide a framework for analyzing effective connectivity in resting state functional MRI with strong *a priori* hypotheses.

*Spectral dynamic causal models*

Although the stochastic models in (10) and their inversion in the time domain provide a useful means to estimate effective connectivity, they also entail the estimation of hidden states. This poses a difficult inverse problem that is computationally demanding; especially when the number of hidden states becomes large. To finesse this problem, a DCM based upon a deterministic model that generates predicted cross-spectra was explored [114, 121]. This scheme provides a constrained inversion of the stochastic model by parameterising the spectral density neuronal fluctuations. This parameterisation also provides an opportunity to compare parameters encoding neuronal fluctuations among groups. The parameterisation of endogenous fluctuations means that the states are no longer probabilistic; hence the inversion scheme is significantly simpler, requiring estimation of only the parameters (and hyperparameters) of the model. The ensuing model inversion in the spectral domain is similar in spirit to previous approaches described in [26, 98, 124]. Put simply, while generalised filtering estimates time-dependent fluctuations in neuronal states producing observed data, spectral DCM simply estimates the time-invariant parameters of their cross spectra. Effectively, this is achieved by replacing the original timeseries with their second-order statistics (i.e., cross spectra). This means that instead of estimating time varying hidden states, we are estimating their covariance. In turn, this means we need to estimate the covariance of the random fluctuations using a scale free (power law) form for the state noise (resp. observation noise) that can be motivated from previous work on neuronal activity [125-127]:

$$g_w(\omega, \theta) = \alpha_w \omega^{-\beta_w}$$

$$g_e(\omega, \theta) = \alpha_e \omega^{-\beta_e}. \tag{11}$$

Here $g_x(\omega) = X(\omega)X(\omega)^\dagger$ represents the complex cross spectra, where $X(\omega)$ is the Fourier transform of the $x(t)$, $\{\alpha, \beta\} \subset \theta$ are the parameters controlling the amplitudes and exponents of the spectral density of the neural fluctuations and $\omega = 2\pi f$ is the angular frequency. This models neuronal noise with generic $1/f^\gamma$ spectra that characterizes fluctuations in systems that are at nonequilibrium steady-state. A linear scaling regime of the spectral density in double logarithmic coordinates – implicit in (11) – is not by itself indicative of a scale free, critical process unless $\gamma$ is less than 1.5 (and the regime scales over several orders of magnitude). For the human EEG, this is generally not the case: above 10 Hz, $\gamma = 2.5$ and above 70 Hz $\gamma$ is usually greater than 3.5, which is consistent with a Poisson process (see [128] and [129]). However, at low frequencies (less than 1.5 Hz) the slope is shallower and it is likely that the amplitude or power envelopes of faster frequencies are scale-free [130, 131] or another heavy-tailed distribution [132]. Using the model parameters, $\theta \supseteq \{A, C, \alpha, \beta\}$, one can simply generate the expected cross spectra as follows:

$$y(t) = \kappa(t) \otimes w(t) + e(t),$$

$$\kappa(t) = \partial_x g \exp(t\, \partial_x f),$$

$$\boldsymbol{g}_{\mathbf{y}}(\omega, \boldsymbol{\theta}) = |\boldsymbol{K}(\omega)|^2 \boldsymbol{g}_{\mathbf{w}}(\omega, \boldsymbol{\theta}) + \boldsymbol{g}_{\mathbf{e}}(\omega, \boldsymbol{\theta}), \tag{12}$$

where $\boldsymbol{K}(\omega)$ is the Fourier transform of the system's (first order) Volterra kernels $\boldsymbol{\kappa}(t)$, which are a function of the Jacobian or effective connectivity (see Figure 3). The unknown quantities $\boldsymbol{\psi} = \{\boldsymbol{\varphi}, \boldsymbol{\theta}, \boldsymbol{\sigma}\}$ of this deterministic model can now be estimated using standard Variational Laplace [133]. The resulting inversion provides the free energy bound on the log evidence $\log p(\boldsymbol{g}_{\mathbf{y}}(\omega)|m)$ and approximate conditional densities $q(\boldsymbol{\psi}) \approx p(\boldsymbol{\psi}|\boldsymbol{g}(\omega), m)$. Here $\boldsymbol{g}_{\mathbf{y}}(\omega)$ represents the predicted cross spectra that can be estimated, for example, using an autoregressive (AR) model.

*An example from ageing*

Finally, in Figure 9 we show an example from recent work on ageing [134] that used spectral DCM. Well-being across the lifespan depends on the preservation of cognitive function. It was hypothesized that successful cognitive ageing is determined by the connectivity within and between large-scale brain networks at rest. Spectral DCM was used to explain the spectral characteristics of resting state fMRI data from 602 healthy adults in a cohort across ages 18-88 (www.cam-can.org). The location of the key cortical regions in each network was identified by spatial ICA, using the Group ICA [120] to extract 20 low-dimensional components. Then the three well-established functional networks: the salience network (SN), dorsal attention network (DAN) and default mode network (DMN) were identified by spatially matching to pre-existing templates [135]. Effective connectivity was assessed within and between these three key large-scale networks; although for brevity we have only included more interesting results for the between network connectivity in this review. In brief, a two-step process is used here, in which ICA is used to identify linearly coherent networks, and then the (potentially non-linear) relationship among these networks is tested within a causal modelling framework using spectral DCM. This approach has been used several times in both task and rest fMRI data [136-138].

Using multiple linear regression, it was found that about 30% of age variance can be predicted ($r = .544$, $p < .001$) by i) increased inhibitory self-connections in SN and DMN networks, ii) decreased effective connectivity from the DAN network to SN network, and iii) increased haemodynamic decay times for all networks (Figure 9, panel B). Subsequently a classical multivariate test (canonical variates analysis) was used to ask to what degree the DCM parameters predict cognitive performance, shown in panel C. For between-network analysis, the corresponding canonical vector suggested that high performance across a range of cognitive tasks (high scores of general intelligence (Cattell), face processing (Benton Faces), memory (story recall), multitasking (Hotel) and response consistency (inverse of response variability of on simple motor task)) was associated with less self-inhibition of the networks and a smaller influence of DMN on SN ($r = 0.447$, $p<0.001$). In other words, about 20% of the variance in performance – across a range of cognitive tasks studied – could be predicted from changes in effective connectivity between networks. To further investigate whether the relationship between cognitive performance and connectivity was age-dependent, moderation analysis was used. It was found that the interaction between age and connectivity values (age x connectivity profile) predicted a significant proportion of variance in cognitive performance, ($T(398) = 3.115$, $p$ (one-tailed) $<0.001$). The direction of the interaction was such that increasing age strengthened the relationship between cognitive and connectivity profiles. This can be seen in panel D, where the relationship between cognitive performance and connectivity profile becomes stronger for older age-groups. This is an interesting study that used spectral DCM to dissociate neuronal from vascular components of the fMRI signal to find age-dependent and behaviorally-relevant differences in resting-state effective connectivity between large-scale brain networks. Taken together, the results suggest that maintaining a healthy resting state connectivity becomes increasingly important for older adults in order to maintain high levels of domain-general cognitive function, and may play a critical role in the mechanisms of healthy cognitive ageing.

*D. Summary*

In summary, both spectral and stochastic DCM furnish estimates of the effective connectivity that underlie intrinsic brain networks. These estimates are based on BOLD data acquired at rest, using different inversion schemes. We suppose that these resting state networks emerge from the dynamical instabilities and critical slowing near transcritical bifurcations. In this setting, neuronal activity is modelled with random differential equations, which can be estimated using stochastic inversion schemes (like generalized filtering in stochastic DCM), or by deterministic schemes modelling observed functional connectivity (specifically the cross spectral densities modelled by spectral DCM).

## V. DISCUSSION

The limitations and challenges of dynamic casual modelling and the implicit scoring of large numbers of models have been addressed in a number of critical reviews (e.g. [139, 140]). Their key conclusions highlight several issues: firstly, although the modelling assumptions underlying DCM are motivated by neuroanatomical and neurophysiological constraints, their plausibility is difficult to fully establish. For example, in DCM for fMRI, physiological details of the neurovascular coupling are potentially important. Many DCMs neglect the potential influence of inhibitory activity on the hemodynamic response, and call on a simplistic account of the metabolic cascade that relates synaptic activity and neuronal metabolism to the vasodilatation. In principle, these are issues that can be resolved using Bayesian model comparison. In other words, if a more complex and complete model is supported by the data, one can always optimize the DCM. Examples of this include recent trends towards more detailed physiological modelling. For example, [141] propose several extensions, such as an adaptive two-state neuronal model that accounts for a wide range of neuronal time courses during stimulation and post-stimulus deactivation; a neurovascular coupling model that links neuronal activity to blood flow in a strictly feedforward fashion; and a balloon model that can account for a vascular uncoupling between blood flow and blood volume due to viscoelastic properties of venous blood vessels.

There are also questions about the robustness of the statistical (approximate Bayesian) inference techniques employed in DCM. For example, it has been argued: i) that the number of parameters and the complexity of the models preclude robust parameter estimation [140, 142], ii) Bayesian model comparison cannot compare DCMs, in the sense that it cannot falsify them and iii) selecting a model based on the model evidence does not ensure that it will generalize. All these concerns stem from frequentist thinking and are dissolved within a Bayesian framework (see [139] for detailed discussion). There are also several – well founded – technical concerns about the variational Bayes (VB) schemes employed in DCM. For example, the objective function based on the free energy functional is prone to local maxima that can result in inconsistent parameter estimations and model comparisons (e.g., across trials or subjects). There are several experimental studies, e.g. [143-147] that have addressed the reproducibility of DCM and provide reassuring experimental validation. There is an issue of overconfidence usually associated with VB schemes, due to the potentially biased inference that results from mean-field and Laplace approximations to the posterior density. This issue has been addressed by simulation studies that compare the results of VB to standard (e.g., Gibb's) sampling methods. The failures of approximate Bayesian inference are usually mitigated by formulating the inversion problem in a way that eschews brittle nonlinearities.

Given issues highlighted above one obvious alternative is to use either 'exact' inference schemes like Markov Chain Monte Carlo or nonparametric methods based on Gaussian processes. Recently both have been explored for inverting Bayesian hierarchical models. For example in [148] Gaussian processes optimization was used for model inversion, [149] explored several gradient-free MCMC schemes (for e.g. random walk based Hasting's sampling, adaptive MCMC sampling and population-based MCMC sampling) and in [150] more robust gradient-based MCMC schemes (for e.g. Hamiltonian and Langevin MCMC sampling) were extensively studied. However, these alternative and promising inference methods are still in an early phase of development and validation phase – and will require exhaustive experimental studies to establish their validity.

Clearly, most of these issues transcend DCM per se and speak to the challenges facing any modelling initiative that has to contend with 'big data' and a large model or hypothesis spaces. These challenges have focussed recent research on contextualizing the inversion of models of single subjects using (empirical or hierarchical) Bayesian models that distinguish between within and between-subject effects on the one hand, and the scoring of large model spaces with techniques like Bayesian model reduction on the other. This is an active research field with developments nearly every month.

Now to conclude, in this review, we used several distinctions to review the history and modelling of macroscopic brain connectivity. We started with the distinction between functional segregation and integration. Within functional integration, we considered the key distinction between functional and effective connectivity and their relationship to underlying models of distributed processing. Within effective connectivity, we have looked at structural and dynamic causal modeling, while highlighting recent advances in the dynamic causal modeling of resting state fMRI data.

We close with a few words on recent large scale projects in neurosciences; for example, the American BRAIN initiative and European Human Brain Project. These initiatives reflect an increasing appreciation of the importance of neuroscience and the challenges of understanding how our brains work. Furthermore, they represent initiatives that exploit remarkable advances in computer science and neuroimaging at many different scales (from the molecular to multi-subject) – and the modelling (and mining) of the resulting data. The experience of the systems neuroscience community, with the big data on offer from neuroimaging, is reflected in this review. This experience highlights the importance of formal models of how data are generated – and the computational schemes used to evaluate and invert these models. We are just embarking on a difficult journey to uncover the governing principles of how brains work and their functional (computational) architectures. Perhaps it is fitting to end on an encouraging quote from Abdus Salam (Nobel prize in physics, 1979) here: "Nature is not economical of structures – only of principles".

## VI. ACKNOWLEDGEMENTS

This work was funded by the Wellcome Trust and CHDI Foundation. We would like to thank our colleagues at Wellcome Trust Centre for Neuroimaging (FIL), on whose work this review is based. We acknowledge very useful comments from reviewers which have greatly helped to improve the readability of this paper.

## VII. FIGURE LEGENDS

**Figure 1**

This schematic depicts a graph with *undirected* edges comprising 10 nodes, where each node can be considered as a neuronal system. We have sketched the evolution of this graph over three time points (under the assumption each node retains a memory of past influences). Nodes 1 and 2 (shown in red) are the nodes whose causal relationship is of interest. The key point of this example is that fluctuations in undirected coupling can induce directed dependencies.

**Figure 2**

No legend

**Figure 3**

This schematic illustrates the relationship among different formulations of dependencies within multivariate timeseries – of the sort used in fMRI. The upper panel illustrates the form of a state-space *model* that comprises differential equations coupling hidden states (first equation) and an observer equation mapping hidden states $\mathbf{x}(t)$ to observed responses $\mathbf{y}(t)$ (second equation). Dynamic causal models are summarised by a Taylor (bilinear) approximation. Crucially, both the motion of hidden states and responses are subject to random fluctuations, also known as state $\mathbf{w}(t)$ and observation $\mathbf{e}(t)$ noise. The form of these fluctuations are modelled in terms of their cross-covariance functions $\mathbf{\Sigma}(t)$ of time *t* or cross-spectral density functions $\mathbf{g}(t)$ of (radial) frequency $\omega$, as shown in the lower equations. Given this state-space model and its parameters $\mathbf{\theta}$ (which include effective connectivity), one can now parameterise a series of *representations* of statistical dependencies among successive responses as shown in the third row. These include convolution and autoregressive formulations shown on the left and right respectively – in either time (pink and orange) or frequency (light green) space. The mapping between these representations rests on the Fourier transform, denoted by (dotted line) and it's inverse. For example, given the equations of motion and observer function of the state-space model, one can compute the convolution kernels that, when applied to state noise, produce the response variables. This allows one to express observed responses in terms of a convolution of hidden fluctuations and observation noise. The Fourier transform of these convolution kernels $\mathbf{\kappa}(t)$ is called a transfer function $\mathbf{K}(t)$. Note that the transfer function in the convolution formulation maps from fluctuations in hidden states to response variables, whereas the directed transfer function in the autoregressive formulation $\mathbf{S}(t)$ maps directly among different response variables. These *representations* can be used to generate second-order statistics or *measures* that summarise the dependencies, as shown in the third row; for example, cross-covariance functions and cross-spectra. The normalised or standardised variants of these measures are shown in the lower row and include the cross-correlation function (in time) or coherence (in frequency). The equations show how various representations can be derived from each other. All variables are either vector or matrix functions of time or frequency. For simplicity, the autoregressive formulations are shown in discrete form for the univariate case (the same algebra applies to the multivariate case but the notation becomes more complicated). Here, $\mathbf{z}(t)$ is a unit normal innovation. Finally, note the Granger causality is only appropriate for bivariate timeseries.

**Figure 4**

Citations rates for resting state resting state fMRI studies. These citations were identified by searching for "fMRI*" and "resting state". *Source: Web of Science*.

**Figure 5**

This schematic illustrates the forward (dynamic causal) model for modelling intrinsic or endogenous fluctuations. The endogenous fluctuations (the state noise), is the driving input to the state-space model of effective connectivity, which is a function of the current neuronal states $\mathbf{x}(t)$, and the connectivity parameters, $\mathbf{\theta}$ that define the underlying structure or functional architecture of the model, and the random fluctuations $\mathbf{w}(t)$. The driving fluctuations cause change in neural activity that, in turn, can be observed using the observer function $h$, after addition of observation noise $\mathbf{e}(t)$. The associated functional connectivity (e.g. cross-covariance function) can be calculated easily from this forward or generative model (see Figure 3) for any given parameters.

Note: The effective connectivity matrix shown is actually a structural connectivity matrix of famous macaque/CoCoMac (as pointed out by one of the reviewers) – we used it here – we use it here as a schematic for effective connectivity.

**Figure 6**

This schematic shows dynamic causal model that embodies the best effective connectivity – identified using Bayesian model inversion (left top panel) – among hidden neuronal states that explains the observed functional connectivity, $\mathbf{\Sigma}(t)$, among haemodynamic responses. This is explanation as possible because the cross-spectra contain all the information about (second-order) statistical dependencies among regional dynamics. Bayesian model inversion furnishes posterior estimates for the parameters of each model and provides the associated log model evidence, in terms of a variational free energy bound. Since the mapping from functional connectivity to effective connectivity is not bijective (there may be many combinations of effective connectivity parameters that induce the same functional connectivity), one can use Bayesian model comparison (top right panel) to score competing models. The model with highest model evidence can then be selected. Alternatively one can use Bayesian model averaging to average across all possible models (bottom left panel).

**Figure 7**

This schematic illustrates the distinction between stochastic and spectral dynamic causal modelling. See text for a detailed description of how these schemes are used to model intrinsic network dynamics.

**Figure 8**

This figure summarizes the effect of subthalamic nucleus deep brain stimulation (STN DBS) on coupling in the motor basal ganglia circuit. Given the marked clinical effect of STN DBS in patients with Parkinson's disease, Kahan et al., (2014) used stochastic DCM to estimate the coupling between key nodes of the basal ganglia network, and whether this coupling is changed by DBS. (A) A network was specified based on human and animal literature, and priors were placed on the nature of the coupling (excitatory or inhibitory), based on the neurochemical systems known to mediate neuronal connections. The literature-based anatomical model of the motor cortico-striato-thalamic loop was further simplified by removing the pallidal nodes and summarizing polysynaptic connections (thick arrows joining the putamen, STN and thalamus). Red arrows constitute excitatory coupling, blue arrows inhibitory coupling. Placing priors on the direction of coupling was enabled using the two-state DCM in the left-hand panel. *The indirect pathway comprised two connections; the striato-STN and STN-thalamus connections (indicated with the dashed grey arrows). (B) Model inversion yielded coupling parameters 'on' and 'off' DBS, demonstrating significant DBS-related changes in extrinsic (between-node) coupling throughout the network. Paired *t*-tests revealed significant differences between extrinsic coupling on and off stimulation. $*P < 0.05$, $**P < 0.001$ (Bonferroni corrected for multiple comparisons). (Upper panel) Cortico-striatal, direct pathway, and thalamo-cortical connections were potentiated by DBS, whereas (lower panel) STN afferents and efferents were attenuated. Note the difference in scale between upper and lower panels. This is because the STN was modeled as a hidden node that was not measured with fMRI. (C) Using a series of regression models, the modulatory effects of DBS on connectivity where shown to predict the clinical improvements seen in the patient cohort. The results are summarized graphically here. [See Kahan et al., (2014) for more details].

**Figure 9**

Regions of interest defined using spatial independent component analysis. **A)** Spatial distribution of three independent components (IC) using group ICA (n = 602) identified as the default mode network (DMN in blue), the dorsal attention network (DAN in red) and salience network (SN in yellow), and the peaks of their corresponding nodes (green circles). Temporal correlation is between (the first eigenvariate of) the ensuing time-series across all nodes and networks. FEF – frontal eye-fields, SPL – superior parietal lobe, vmPFC – ventromedial prefrontal cortex, PCC – parietal cingulated cortex, IPL – intraparietal lobe, dACC – dorsal anterior cingulated cortex, AI – anterior insula, r – right, l – left, all – SVD of all voxels, across all nodes, within a given network. **B)** Coefficients for how well a) effective connectivity (white), b) neuronal (green) and c) haemodynamic (red) parameters predict age. DCM parameters with bars (95% confidence intervals) that exclude zero are considered as significant predictors. **C)** Between network canonical variates analysis. Heliograph of variate loadings for the first canonical variate, where the relative sizes of correlations are indicated by the relative length of the bars (the dark is positive, the white is negative). These reflect the statistical relationship between variables of effective connectivity (Connectivity profile) and cognitive performance (Cognitive profile) (r = .440, p < .001). Variables with low contribution (r < .3) are shown as bars with dashed outline. Half-maximum strength of correlation is indicated by the dashed rings (outer is r = +0.5, inner is r = -0.5). **D)** Corresponding bivariate canonical correlation for three age groups. The relationships between connectivity and cognitive profiles are more pronounced for older subjects, suggesting that performance in older adults reflect a preserved connectivity.

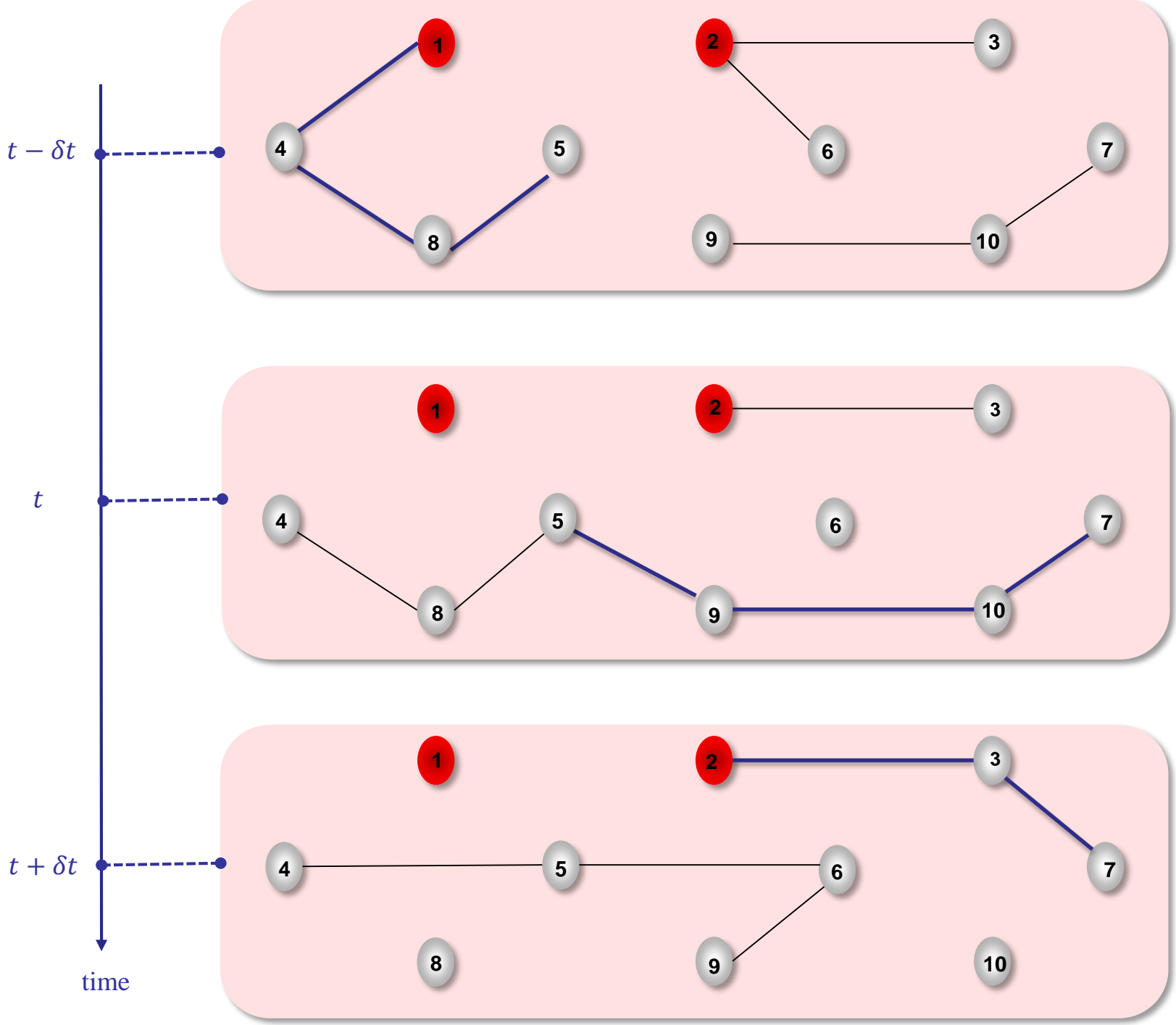

**Fig 1**

**Fig 2**

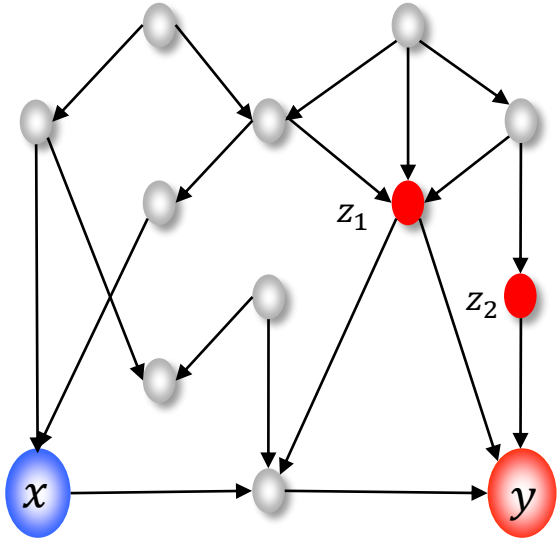

Step: 1

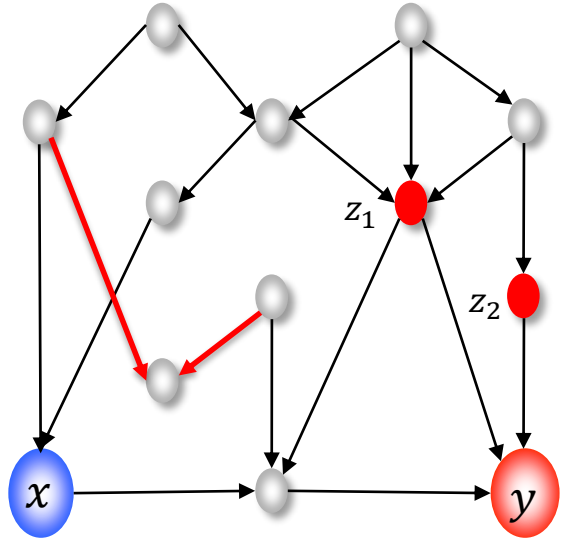

Step: 2

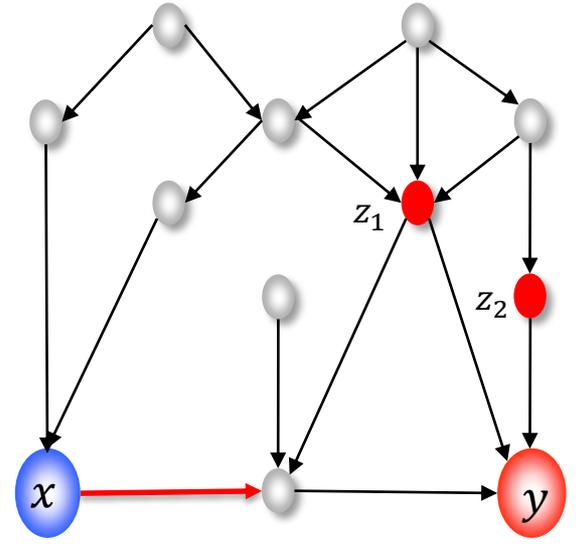

Step: 3

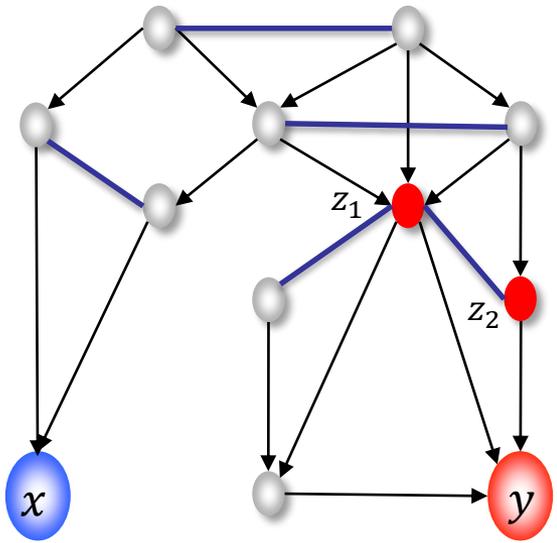

Step: 4

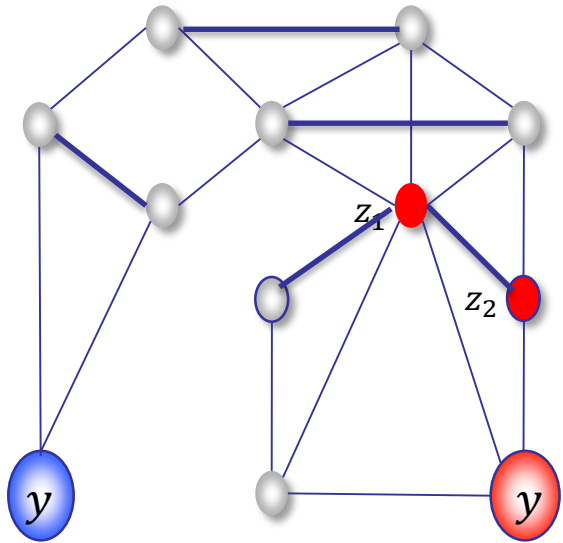

Step: 5

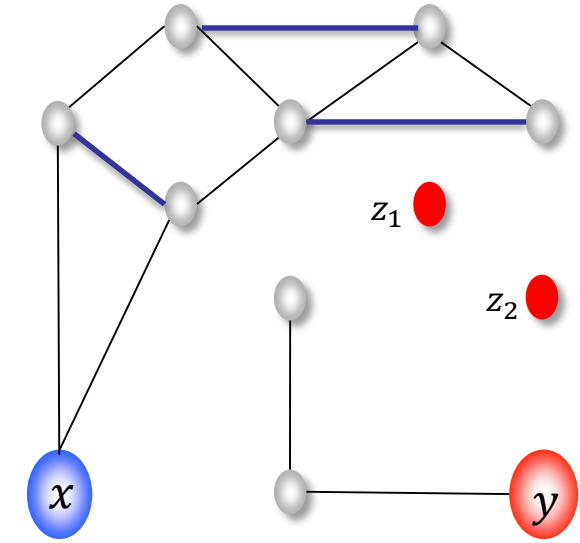

Step: 6

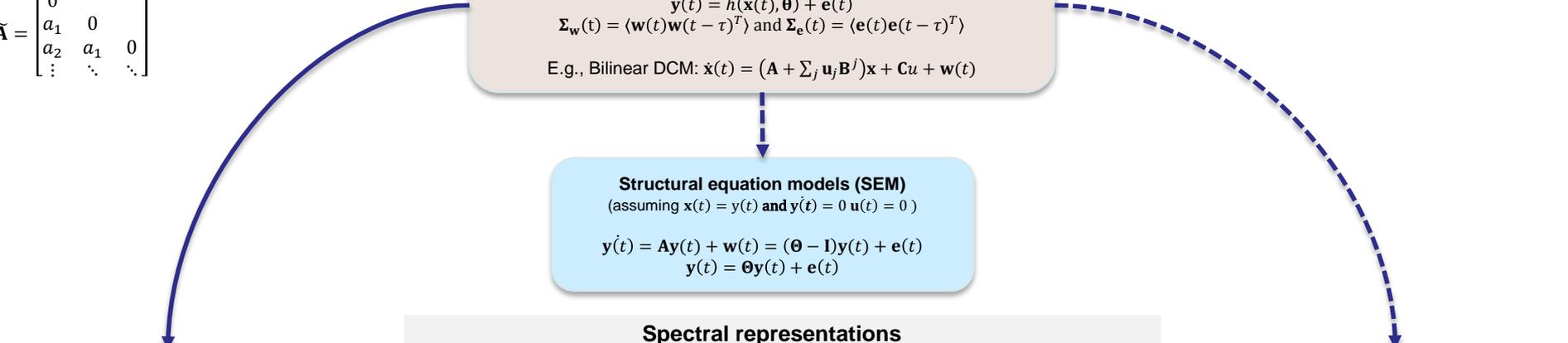

Fig 3

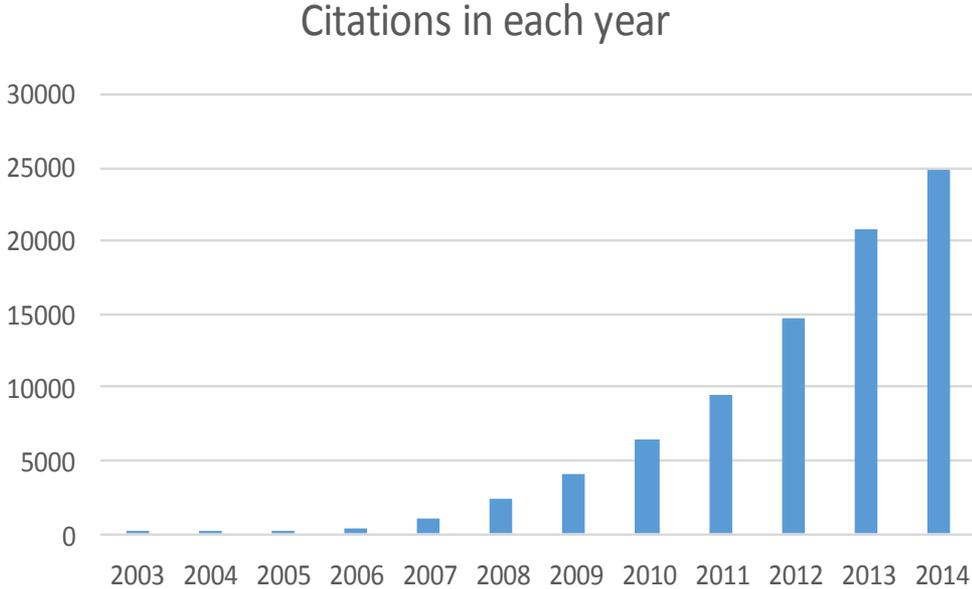

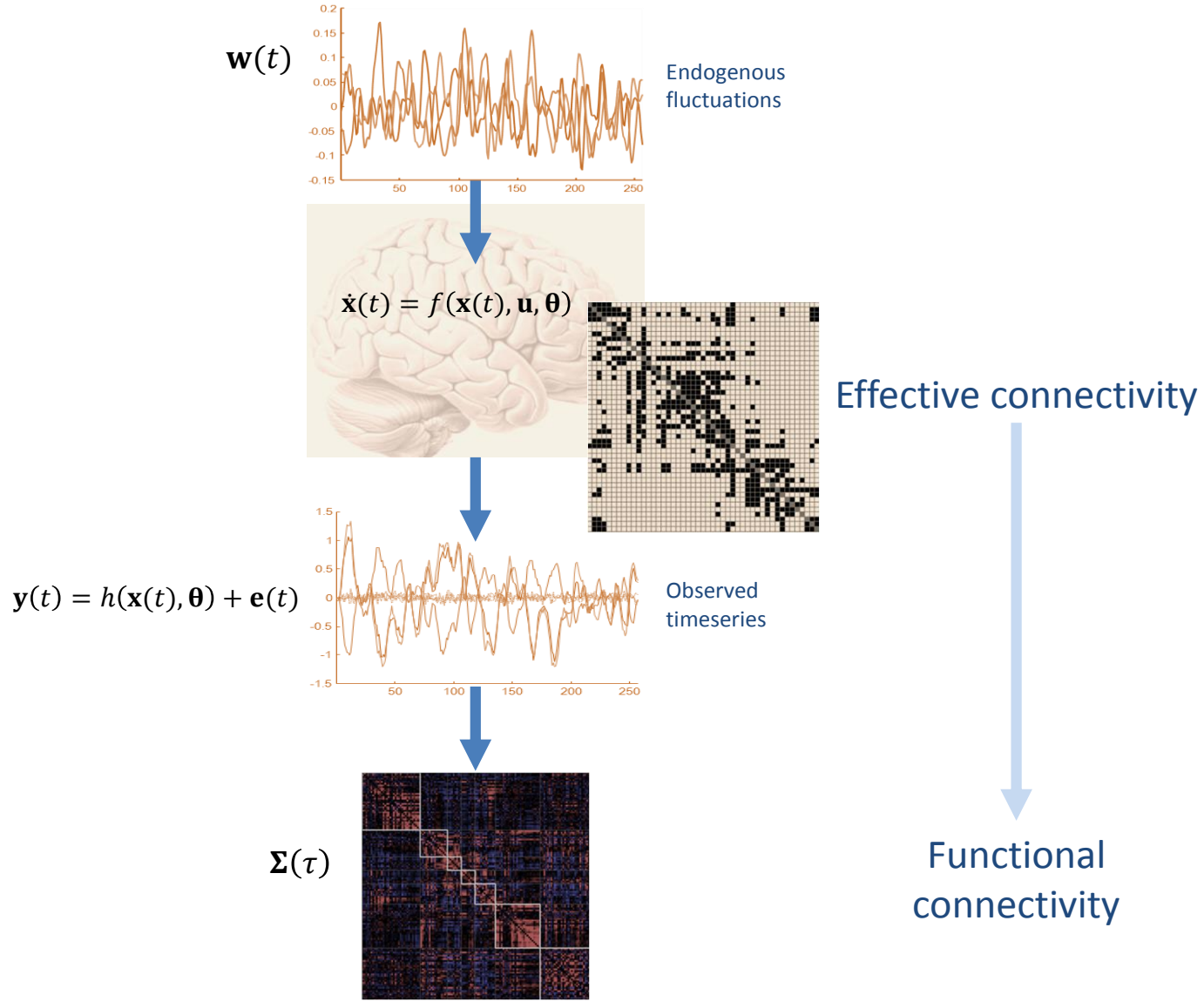

**Fig 6**

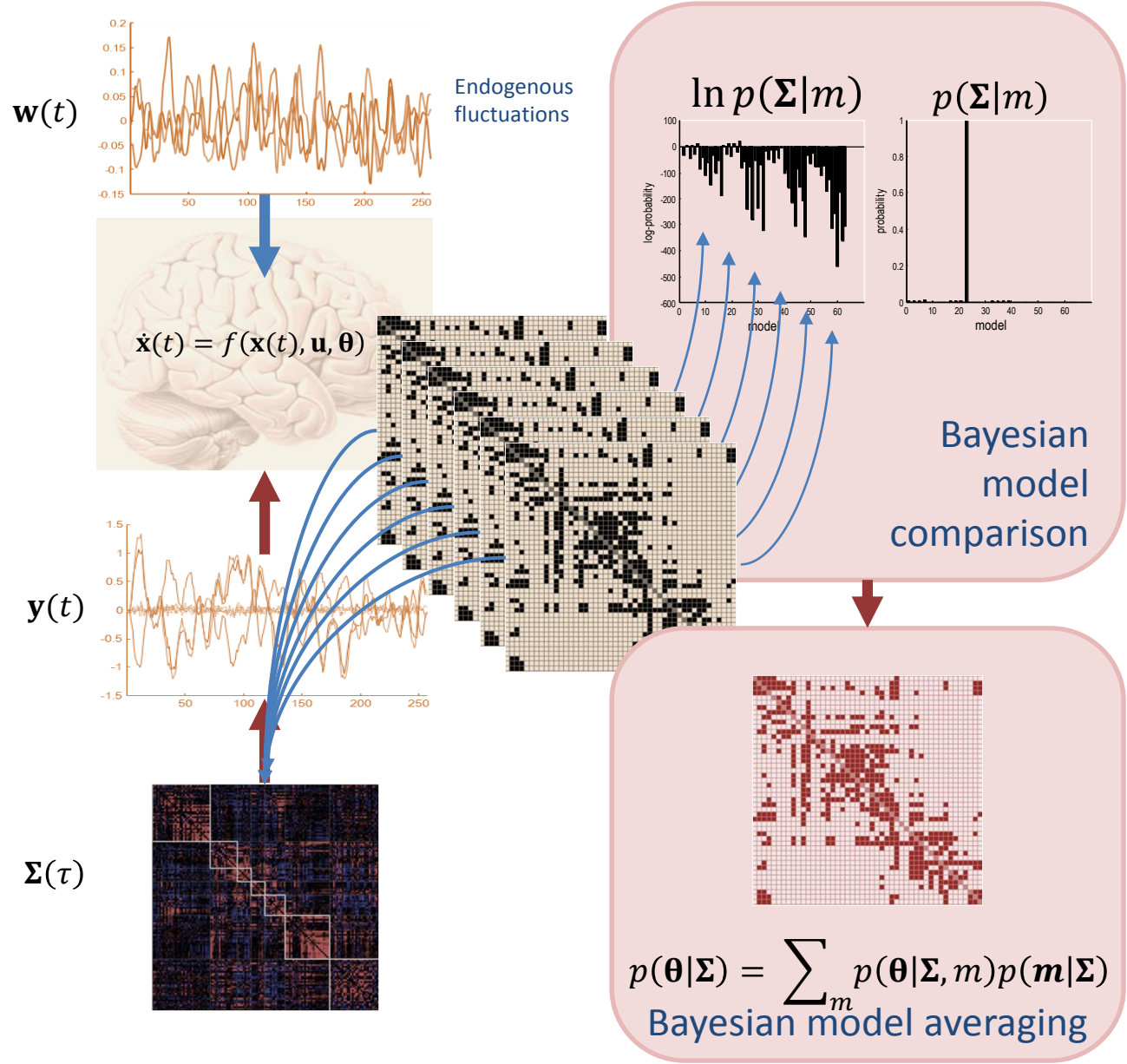

## Stochastic DCM

## Spectral DCM

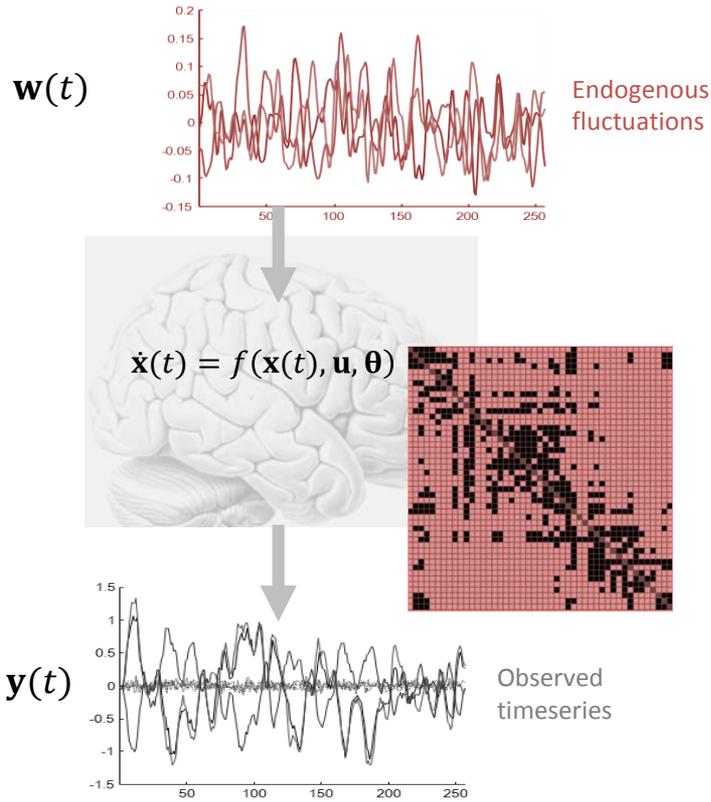
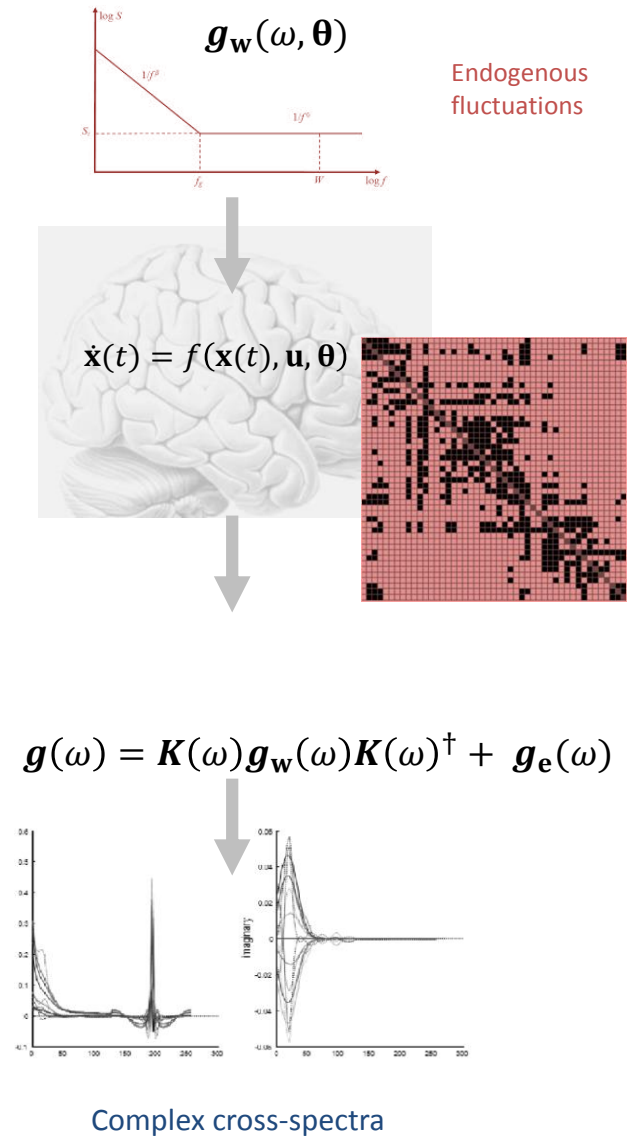

**Fig 8**

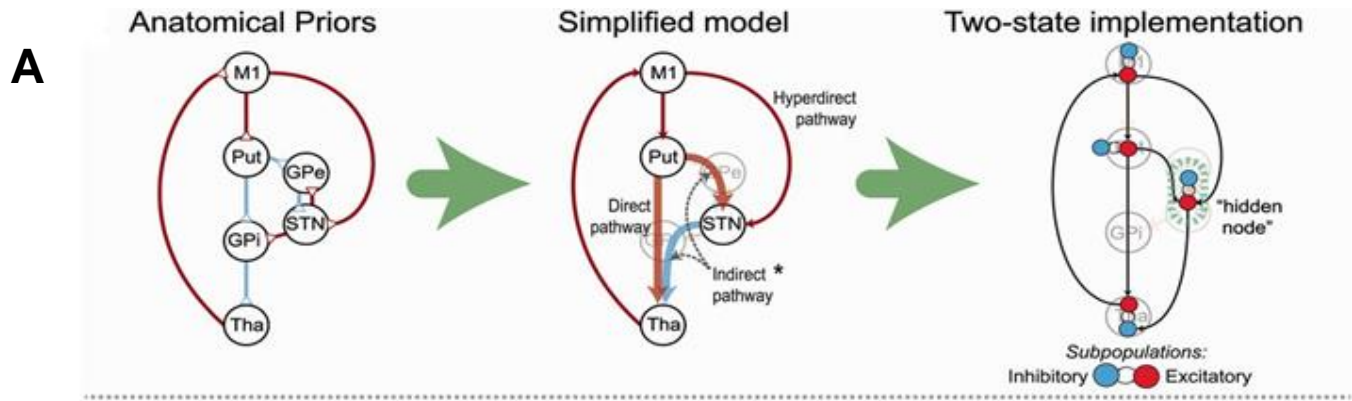
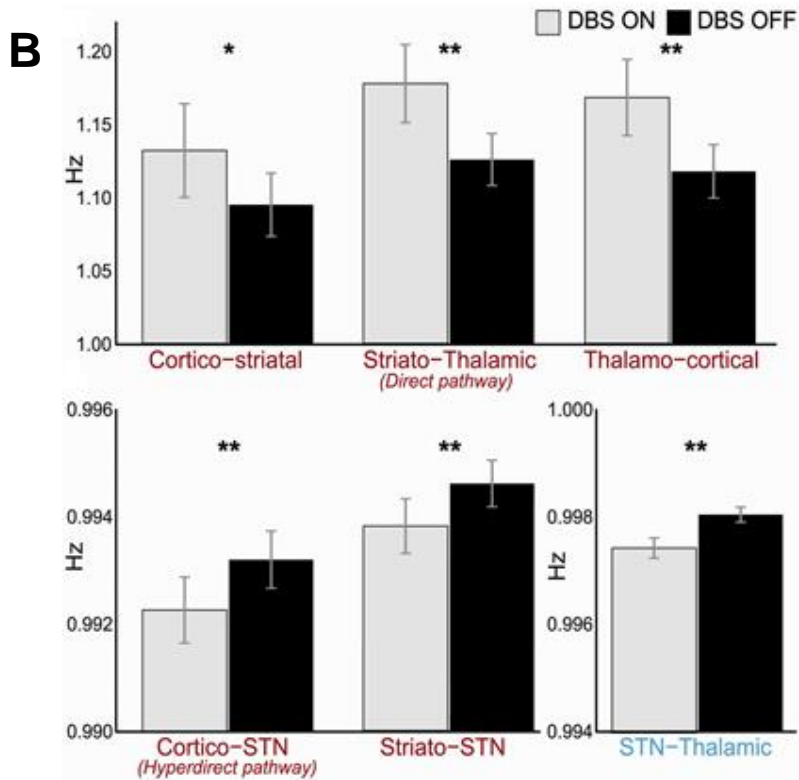
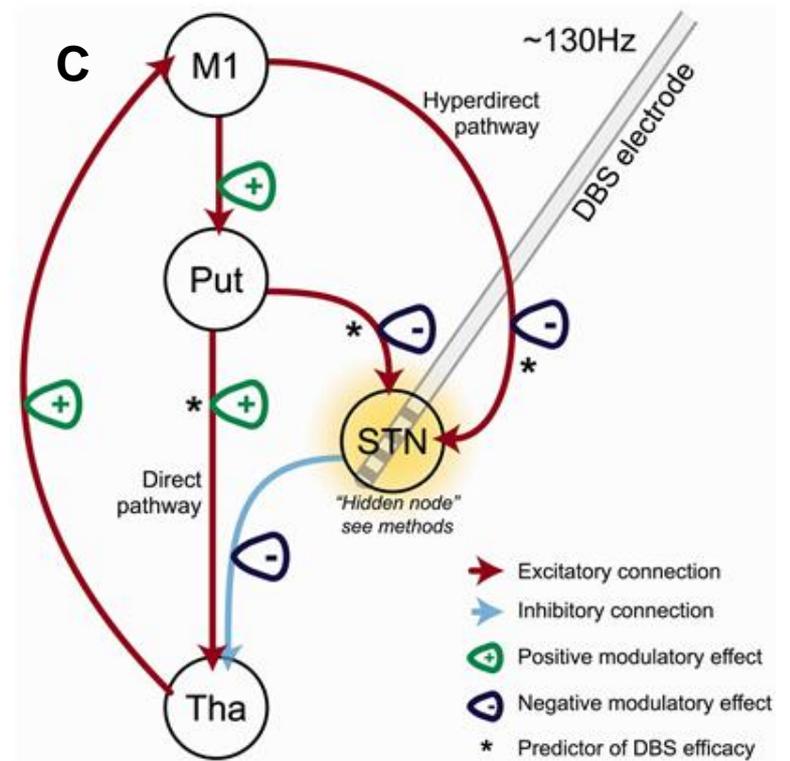

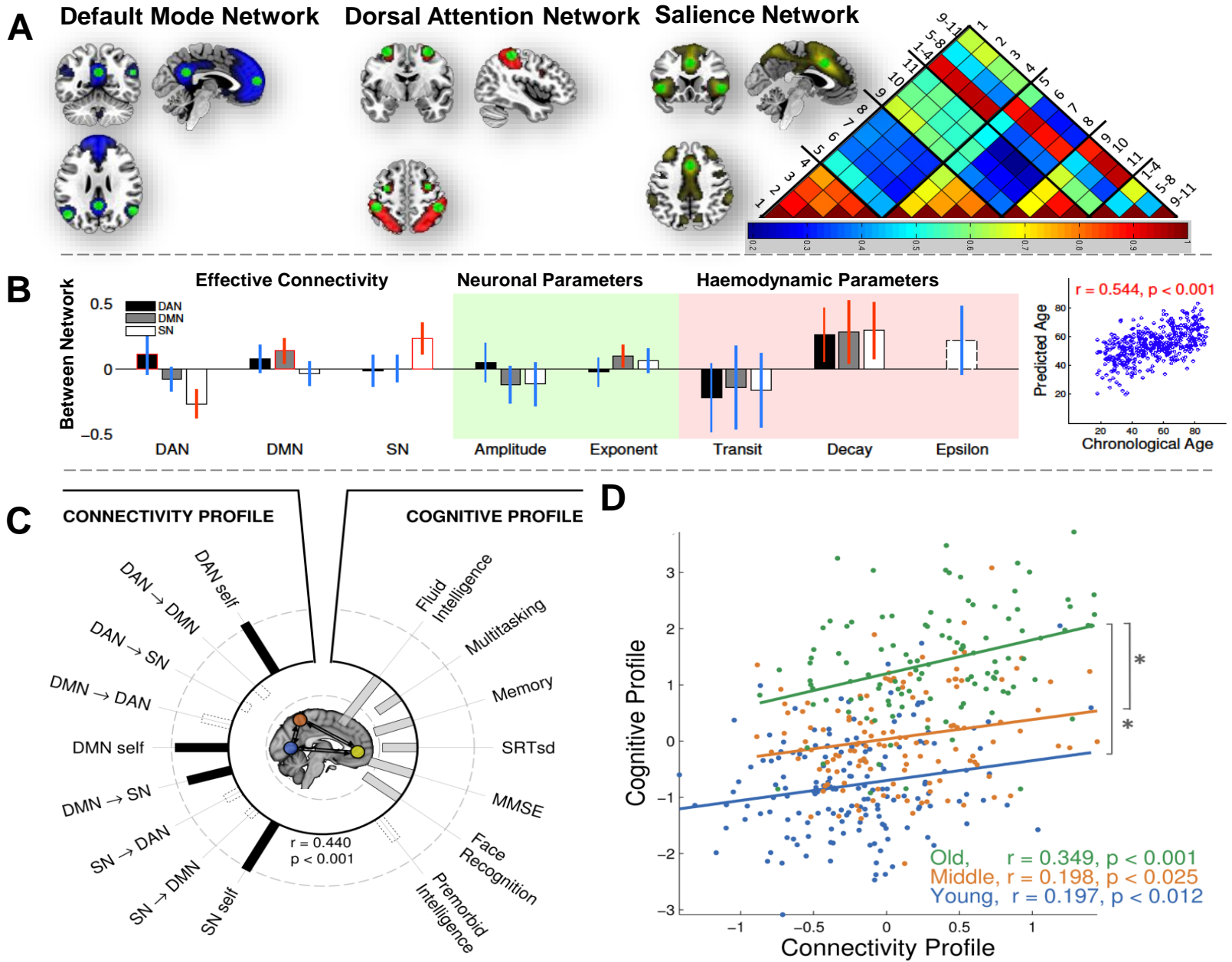